\documentclass[oneside,twocolumn,aps,pra,preprintnumbers,bibnotes10pt,superscriptaddress,amsmath,amssymb]{revtex4-2}

\usepackage[utf8]{inputenc}
\usepackage{graphicx} 
\usepackage{subcaption}
\usepackage[usenames,dvipsnames]{xcolor}
\usepackage{epstopdf}
\usepackage{amsmath}
\usepackage{braket}
\usepackage{amsthm}
\usepackage{amssymb}
\usepackage{amsfonts}
\usepackage{txfonts}
\usepackage{ulem}
\usepackage{mathtools}
\usepackage{bm}
\usepackage{hyperref}
\usepackage{lipsum}
\usepackage{natbib}
\usepackage{comment}
\usepackage{physics}
\usepackage{cancel}

\usepackage{lipsum}  

\usepackage{tikz}
\usepackage{tikzsymbols}
\usetikzlibrary{quantikz2,calc,arrows,chains,matrix,positioning,scopes}

\usepackage{letltxmacro}
\LetLtxMacro{\originaleqref}{\eqref}
\renewcommand{\eqref}{Eq.~\originaleqref}

\begin{document}

\title{Impact of quantum coherence on the dynamics and thermodynamics of quenched free fermions coupled to a localized defect}

\author{Beatrice Donelli}
\affiliation{Istituto Nazionale di Ottica del Consiglio Nazionale delle Ricerche (CNR-INO), Largo Enrico Fermi 6, I-50125 Firenze, Italy.}
\affiliation{European Laboratory for Non-linear Spectroscopy, Università di Firenze, I-50019 Sesto Fiorentino, Italy.}

\author{Gabriele De Chiara}
\affiliation{Física Teòrica: Informació i Fenòmens Quàntics, Departament de Física, Universitat Autònoma de Barcelona, 08193 Bellaterra, Spain.}
\affiliation{Centre for Quantum Materials and Technology, School of Mathematics and Physics, Queen’s University Belfast, Belfast BT7 1NN, United Kingdom.}

\author{Francesco Scazza}
\affiliation{Department of Physics, University of Trieste, 34127 Trieste, Italy}
\affiliation{Istituto Nazionale di Ottica del Consiglio Nazionale delle Ricerche (CNR-INO), 34149 Trieste, Italy}
    
\author{Stefano Gherardini}
\affiliation{Istituto Nazionale di Ottica del Consiglio Nazionale delle Ricerche (CNR-INO), Largo Enrico Fermi 6, I-50125 Firenze, Italy.}
\affiliation{European Laboratory for Non-linear Spectroscopy, Università di Firenze, I-50019 Sesto Fiorentino, Italy.}

\begin{abstract}
We investigate the non-equilibrium quantum dynamics and thermodynamics of free fermions suddenly coupled to a localized defect in a one-dimensional harmonic trap. This setup realizes a quantum quench transformation that gives rise to the orthogonalization of the system's wave-function as an effect of the localized perturbation. Using the Loschmidt echo and the Kirkwood-Dirac quasiprobability (KDQ) distribution of the work done by the defect, we quantify the extent and rate of the orthogonalization dynamics. In particular, we show that initializing the system in a coherent superpositions of energy eigenstates leads to non-classical features, such as Wigner function's negativity and non-positivity of the work KDQ distribution. Starting from simple single-particle superpositions and then progressing with coherent and cat states of few-body fermionic systems, we uncover how quantum coherence and few-body correlations shape the out-of-equilibrium response due to the presence of the defect.
\end{abstract}

\maketitle

\section{Introduction}

The orthogonalization of a fermionic many-body wave-function in the presence of a localized impurity is a paradigmatic phenomenon in quantum many-body physics. This effect, first theorized in the seminal work of Anderson~\cite{AndersonPRL1967} and typically coined {\it orthogonality catastrophe}, shows how a single static impurity can drive the ground state of a fermionic system to an orthogonal state in the thermodynamic limit of many particles. The orthogonality catastrophe has many implications ranging from condensed matter to ultracold atom physics, such as divergences in the optical conductivity of interband valence-conduction transitions~\cite{MahanPR1967} and universal impurity dynamics~\cite{Knap2012,Cetina2016}.

A powerful tool to capture the effects of the orthogonality catastrophe is the Loschmidt echo (LE)~\cite{PeresPRA84,JalabertPRL2001,CucchiettiPRL2003,SilvaPRL2008,LostaglioQuantum2023}, which quantifies the overlap between the eigenstates of two different Hamiltonians, including or excluding the interaction term that models the sudden coupling with the defect~\cite{goold_orthogonality_2011}. The LE captures how the system dynamically ``remembers'' its initial configuration and reveals the extent of the state orthogonalization over time. The decay of the LE encodes information about the dimensionality of the fermionic system and the nature of the impurity. In particular, performing the Fourier transform of the LE, $\nu(t)$, at time $t$ gives further information, as originally provided by the dynamical theory of the orthogonality catastrophe developed by Nozi\`eres and De Dominicis~\cite{NozieresPR1969}. 
In fact, as outlined also in Ref.~\cite{goold_orthogonality_2011}, if the Fourier transform of $\nu(t)$ decays as a power law instead of an exponential function or if it is asymmetric with respect to $\omega$, with $\omega$ being the frequency-variable (in particular, the asymmetry between values at $\omega$ and $-\omega$), then the fermionic system lies in an out-of-equilibrium regime.

So far, orthogonalization has been thought as an emergent macroscopic effect; however, it is becoming apparent that finite-size effects imply non-trivial quantum dynamical signatures of this phenomenon. Understanding these dynamics, particularly in the non-equilibrium regime following a quantum quench transformation~\cite{SilvaPRL2008,Fusco2014,AlbaPNAS2017,santini_work_2023,GherardiniTutorial}, is essential for exploring the interplay between quantum coherence and/or correlations and irreversibility in many-body quantum systems~\cite{YoshimuraPRE2025}.

In this paper, we investigate the quantum dynamics and thermodynamics of a few-fermion system that is suddenly coupled to a localized defect. The system is trapped in a one-dimensional harmonic potential, and the defect is introduced as a delta-function potential at the center of the trap ~\cite{Jaksch2005,Recati2005,Knap2012,Cetina2016,paper_short}. Our aim is to understand how a local, sudden quench of this defect affects the out-of-equilibrium evolution of the fermionic quantum state.

We find that the presence of a delta-function perturbation generally induces negativity in the Wigner function of the perturbed system's state, even globally across the entire position-momentum ($x-p$) space. This fact can hold irrespectively of the initial state. However, more intriguingly, quantum effects emerge when choosing initial states that do not commute with either the unperturbed or the perturbed Hamiltonians. Indeed, initializing the system in a coherent superpositions of energy eigenstates, we observe the emergence of non-classical features such as non-positivity in the Kirkwood-Dirac quasiprobability (KDQ) distribution~\cite{yunger2018quasiprobability,ArvidssonShukurJPA2021,DeBievrePRL2021,LostaglioQuantum2023,BudiyonoPRAquantifying,wagner2023quantum,GherardiniTutorial,ArvidssonShukur2024review,hernandez2024Interfero} of the work done by the defect on the fermionic system. We are going to show that the Fourier transform of this KDQ distribution is exactly equal to the LE.

To clearly single-out the non-positivity of the work KDQ distribution, we initially focus on a single particle initialized in the superposition of two states --- namely, a qubit --- which admits a representation on the Bloch sphere. In particular, we consider the superpositions of two low-energy eigenstates of the unperturbed Hamiltonian. Despite its simplicity, this initial configuration captures essential features of how the presence of quantum coherence in the initial state significantly alters the dynamics of the particle, with emerging non-classical behaviours. Furthermore, this setting provides a clear interpretation of the role played by phase and amplitude terms of the initial superposition state in controlling the orthogonalization of the particle's state. Afterwards, we extend the analysis to coherent and cat states, which induce richer interference effects, and finally to case-studies with two fermions where we explore how anti-symmetrization and quantum correlations modulate the response of the fermionic system to the presence of the defect.

In our analysis we quantify the average work done by the defect on the fermionic system $\langle w \rangle$. Given that the defect acts as a localized perturbation driving the system out-of-equilibrium, we find that a larger $\langle w \rangle$ is generally accompanied by a faster orthogonalization time of the system's wave-function. This observation suggests a direct link between energetic contributions from localized perturbations and the quantum speed limit~\cite{Fogarty2020,paper_short}, providing a thermodynamic interpretation of the mechanisms that accelerate the departure of the system from its initial configuration.

The paper is structured as follows. In Sec.~\ref{sec:model}, we introduce the Hamiltonian model that describes the interaction between a fermionic system and a single, localized defect, modelled as a delta-perturbation. In Sec.~\ref{sec:Wigner} we discuss the impact of the delta-perturbation on the Wigner function of the system's state. On the other hand, in Sec.~\ref{sec:KDQ_distribution} we report the definition and properties of the KDQ distribution that defines the distribution of work done by the defect, while in Sec.~\ref{sec:LE_Pw} we formally derive the LE for the considered setting, and we show how to measure the real and imaginary part of the LE via an interferometric scheme. In Sec.~\ref{sec:overlap_many-body_state} we derive analytical expressions for the overlaps between the eigenstates of the unperturbed and perturbed Hamiltonian of the system, before and after the quench transformation respectively. We do this both in the limit of strong and weak perturbation strength. The average and variance of the work distribution done by the defect on the quantum system is investigated in Sec.~\ref{sec:Average_variance_work}, while in Sec.~\ref{sec:single_particle} we present detailed case-studies focusing first on a single-particle system prepared, respectively, in qubit, coherent and cat states. Then, in Sec.~\ref{sec:two_fermions}, we extend the analysis to a two-fermion system, where many-body correlations and anti-symmetrization effects become important. Finally, in Sec.~\ref{sec:conclusion} we draw our conclusions and discuss possible future perspectives.

\section{Model}\label{sec:model}

Let us consider a single, localized defect coupled to a free-fermion gas. In our description, the defect (D) is represented by a localized motionless neutral atom. For experimental purposes, the defect can be described as a two-level quantum system, with $\ket{0}_D$ and $\ket{1}_D$ denoting the ground and excited states of its free Hamiltonian $\hat{H}_D$, respectively. Calling $\hbar\omega_D$ the energy difference between these two levels, the Hamiltonian of the defect is 
$\hat{H}_D = \frac{\hbar\omega_D}{2} \hat{\sigma}^z_D$ with $\hat{\sigma}^z_D = |0\rangle_D \langle 0| - |1\rangle_D \langle 1|$. The system (S), in which this defect is suddenly immersed via a quench dynamics, is a fermionic gas in a one-dimensional harmonic trap with Hamiltonian
\begin{equation}\label{H_perturbation}
    \hat{H}_S = \int \hat{\Upsilon}^{\dagger}(x) \left( -\frac{\hbar^2}{2m} \frac{d^2}{dx^2} + V_{\rm h.o.} \right) \hat{\Upsilon}(x) \ dx,
\end{equation}
where $\hat{\Upsilon}(x)$ denotes the fermionic field operator, $V_{\rm h.o.}=\frac{1}{2}m\omega^2x^2$ is the harmonic trap potential with frequency $\omega$, $m$ is the fermionic mass and $\hbar$ is the reduced Planck constant. Considering the defect to be strongly localized at $x=0$, we model its interaction with the fermionic system in the following way: the two internal levels of the defect correspond to the absence or presence of an interaction with the system. In the latter case, valid at sufficiently low temperatures, the interaction potential can be effectively approximated by a contact term proportional to the Dirac delta function $\delta(\cdot)$:
\begin{equation}
    \hat{H}_I = k \int \hat{\Upsilon}^{\dagger}(x) \delta(x) \, \hat{\Upsilon}(x) \ dx,
\end{equation}
where $\delta(\cdot)$ is the Dirac delta function and $k$ is the strength of the interaction.

In order to show the effects of a delta-function potential as the interaction Hamiltonian (also denoted as delta-perturbation throughout the text), we consider a one-dimensional harmonic trap of frequency $\omega$ with a Dirac $\delta$ potential of strength $k$ in the middle of the trap, following \cite{busch_low-density_2003}. The corresponding dimensionless single-particle Hamiltonian is 
\begin{equation}\label{HamiltonianDeltaPotential}
    \hat{h} = - \frac{d^2}{d\Tilde{x}^2} + \frac{1}{4} \Tilde{x}^2 + \Tilde{k} \delta(\Tilde{x}),
\end{equation}
where we have used the following dimensionless variables:
\begin{equation}
    \Tilde{x} = \frac{x}{a}, \quad a=\sqrt{\frac{\hbar}{2m\omega}}, \quad \Tilde{t} = t\omega, \quad \Tilde{k} = \frac{k a}{\hbar \omega} \,. 
\end{equation}
When $\Tilde{k}=0$, the single-particle ground state of \eqref{HamiltonianDeltaPotential} reduces to that of the unperturbed quantum harmonic oscillator (QHO). In the opposite limit, $\Tilde{k}\rightarrow+\infty$, the ground state approaches the first excited state of the QHO. This occurs because all odd-parity eigenfunctions of the Hamiltonian $\hat{h}$ have a node at $\Tilde{x}=0$, where the delta-function is located. 
As a result, these states are not affected by the presence of the defect. Consequently, the ground state energy of \eqref{HamiltonianDeltaPotential} takes the dimensionless value $\Tilde{E}=E/\hbar\omega=1/2$ at $\Tilde{k}=0$, and asymptotically tends to $\Tilde{E}=3/2$ as $\Tilde{k}\rightarrow+\infty$. Hereafter, all variables will be dimensionless, and the tilde notation will be omitted unless necessary.

Examining the eigenenergies of the single-particle Hamiltonian, one finds that the energies of all odd eigenstates remain unchanged by the delta-perturbation, coinciding with those of the unperturbed system. On the other hand, the energy of the even eigenstates of the perturbed Hamiltonian tends to the energy of the eigenstates with subsequent number of levels of the unperturbed Hamiltonian, in the limit of $k \rightarrow +\infty$. The analysis for the many-body wave-function of $N$ non-interacting fermions, with and without the presence of the defect, is carried out in Sec.~\ref{sec:overlap_many-body_state}.   

\section{Delta-perturbation implies Wigner function negativity}\label{sec:Wigner}

At the level of the system wave-function, the presence of a localized defect modelled as a delta-function potential introduces a non-trivial quantum dynamical effect, which can be analysed by examining the Wigner function in phase-space after the quench transformation.

The effects due to the delta-perturbation are quite drastic as it necessarily induces negativity in the Wigner function of the system's state, independently on both the initial state and the number of particles of the fermionic system. Thus, even when the system is initialized in a state with a positive Wigner distribution, such as the unperturbed ground state or a coherent state, Wigner function's negativity arises. As is well-known in quantum optics~\cite{Wigner1932,Gerry2004,Loudon2000}, this negativity is a hallmark of the presence of quantum coherence and interference, and provides a clear evidence for the onset of non-classicality resulting from a sudden perturbation. 

\begin{figure}
    \centering
    \includegraphics[width=\linewidth]{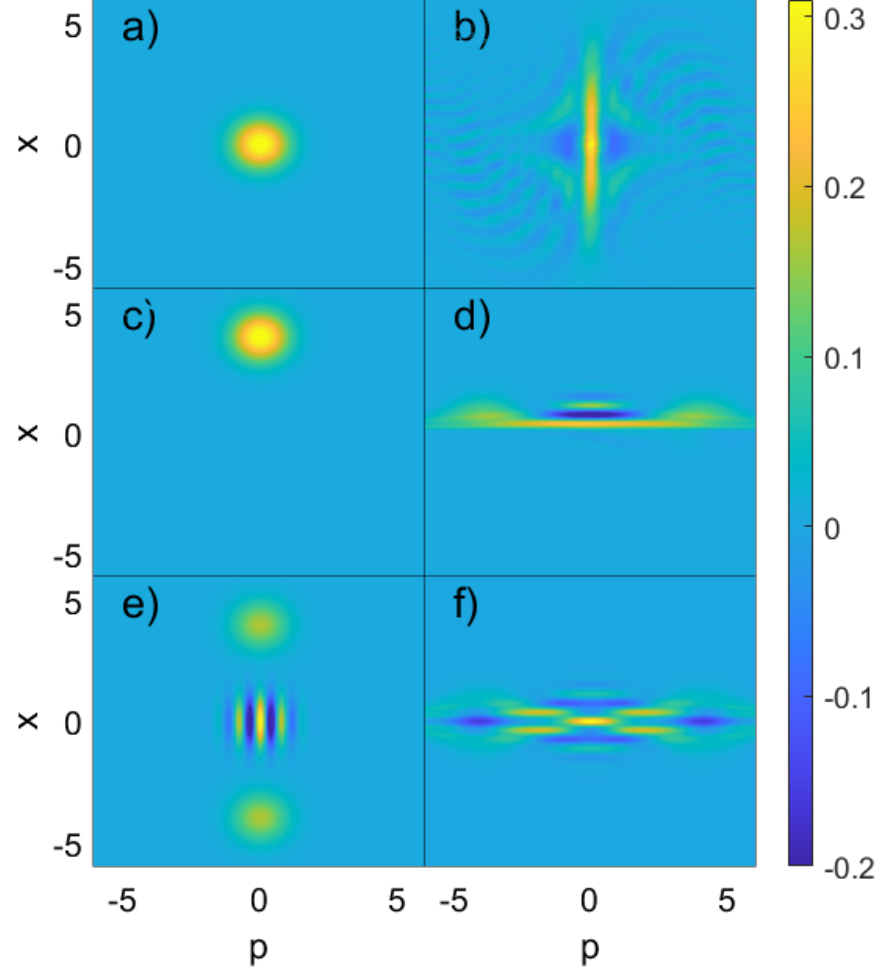}
    \caption{Contour plot in the position-momentum ($x-p$) space, showing the value (identified by the colorbar) of the Wigner functions of the state before (panels at the left) and after (panels at the right) the quench transformation at $t=\pi/2$ that makes interact the system with a localized defect. The panels differ from the choice of the initial state of the system: (a)-(b) ground state of the QHO; (c)-(d) coherent state; (e)-(f) cat state.}
    \label{fig:Wigner}
\end{figure}

In Fig.~\ref{fig:Wigner}, for illustrative purpose, we consider a single particle subjected to the delta-perturbation and observe the generation of negativity for different initial states. In particular, three representative cases are shown in the rows of Fig.~\ref{fig:Wigner}, from top to bottom: (i) the ground state of the QHO, $\ket{\psi_0}=\ket{0}$; (ii) a coherent state built on the basis defined by the eigenstates of the QHO's Hamiltonian, representing in our setting the unperturbed Hamiltonian before the quantum transformation:
\begin{equation}
    |\psi_0\rangle = e^{-|\xi|^2/2} \sum_{n=0}^{+\infty} \frac{ \xi^n }{\sqrt{n!}} \ket{n},
\end{equation}
with $\xi\in\mathbb{C}$; (iii) a cat state defined as the symmetric superposition of two coherent states with opposite phases: $\ket{\psi_0} \propto \ket{\xi} + \ket{-\xi}$.

Under evolution with the unperturbed Hamiltonian, the Wigner function rotates in phase space about the origin. In contrast, the presence of a delta-perturbation due to the quench transformation manifests as a sudden cut of the Wigner function at $x=0$. The ground state, shown at $t=0$ in panel (a) of Fig.~\ref{fig:Wigner}, is effectively split in half by the delta-perturbation, developing a long stripe that subsequently rotates and elongates along the $x$-axis at $t=\pi/2$ [see panel (b)]. The Wigner function of the coherent state with $\xi=4$ initially lies on the left side of the delta-perturbation and remains confined to that region, as shown in panels (c)-(d). As the state evolves, the Wigner function rotates and is reflected by the delta-perturbation, effectively compresses against it. During and after the perturbation, the Wigner function undergoes an inversion of the momentum $p$ from positive to negative, before eventually returning to its original configuration. The cat state is given by the superposition of two coherent states localized on either side of the delta-perturbation [panel (e)], and exhibits interference fringes at $x=0$ where the delta-perturbation is located. This symmetric configuration causes the dynamics to be symmetric under a rotation of $\pi$ and produces the more interesting pattern shown in panel (f) at $t=\pi/2$.

Having established that the delta-perturbation induces negativity in the Wigner function of the state of the perturbed system, we are interested next in quantifying how this reflects in measures of non-classicality involving two-time quantum correlators. To do this, for convenience, we will work with a Kirkwood-Dirac representation of joint probability distribution in the quantum regime~\cite{LostaglioQuantum2023,GherardiniTutorial,ArvidssonShukur2024review}.

\section{Kirkwood-Dirac quasiprobability distribution}\label{sec:KDQ_distribution}

KDQs describe the multi-time statistics associated to outcomes' tuples originated by a sequence of quantum measurements, under the (possible) non-commutativity of the initial density operator $\hat{\rho}$ with the measurement observables~\cite{GherardiniTutorial}. The incompatibility of the measurement observables --- in our case $\hat{O}_1(t_1)$ and $\hat{O}_2(t_2)$ that we assume to measure at times $t_1$ and $t_2$ with $t_1<t_2$ --- can entail these quasiprobabilities to not longer obey all Kolmogorov's axioms of classical probability, though they sum to $1$, are linear in $\hat{\rho}$ and returns the correct marginals distributions (summing over the eigenvalues of $\hat{O}_1(t_1)$ and $\hat{O}_2(t_2)$ respectively). The fact that a KDQ may not respect the classical theory of probability, thus admitting a non-classical trait, reflects in being a negative real number or even a complex one. We can refer to each of these circumstances as a loss of positivity of the multi-time measurement outcomes' distribution. In the following, in relation to two-time statistics, we will consider the KDQ
\begin{equation}\label{eq:def_KDQ}
    q_{s_1,s_2} = {\rm Tr}\left[ \hat{\Pi}_{s_2}^H(t_2) \hat{\Pi}_{s_1}(t_1) \hat{\rho} \right],
\end{equation}
where we consider that $\hat{\rho}$ is initialized at time $t=t_1$, and $o_{s_1}, o_{s_2}$ are the outcomes obtained from measuring $\hat{O}_1, \hat{O}_2$. In \eqref{eq:def_KDQ}, the superscript $H$ denotes the evolution of a given observable in the Heisenberg representation, and $\hat{\Pi}_{s_i}$ ($i=1,2$) are the projectors of the measurement operators $\hat{O}_1(t_1)$ and $\hat{O}_2(t_2)$ that can be obtained from their spectral decomposition: 
\begin{equation}
    \hat{O}_{i}(t_{i}) = \sum_{s_{i}} o_{s_{i}}(t_{i}) \hat{\Pi}_{s_{i}}(t_{i})\,.
\end{equation}
If we introduce $\Delta o_{s_1,s_2} = o_{s_2}(t_2) - o_{s_1}(t_1)$ as a generic difference between the measurement outcomes, then the probability distribution of $\Delta o$ in terms of the KDQ of \eqref{eq:def_KDQ} is expressed as
\begin{equation}
    P[\Delta o] =  \sum_{s_1,s_2} q_{s_1,s_2} \delta(\Delta o - \Delta o_{s_1,s_2}) \,. 
\end{equation}
The same information carried by the distribution $P[\Delta o]$ is contained in its complex Fourier transform, which is known as the characteristic function of $\Delta o$. Formally, it reads as
\begin{align}
    \mathcal{G}(u) &\equiv \int_{-\infty}^{+\infty} P[\Delta o] e^{i u \Delta o} d\Delta o = \nonumber \\
    &= \sum_{s_1,s_2} q_{s_1,s_2} e^{i u \Delta o_{s_1,s_2}} = \nonumber \\
    &= {\rm Tr}\left[ e^{-i u \hat{O}_1(t_1)} \, \hat{\rho} \, \hat{U}^\dagger(t_2,t_1) e^{i u \hat{O}^H_2(t_2)} \hat{U}(t_2,t_1) \right],
\end{align}
where $u$ is the complex number with respect to which the Fourier transform is performed, and $\hat{U}(t_2,t_1)\equiv e^{-i(\hat{H}_S+\hat{H}_I)(t_2-t_1)}$ is the unitary evolution operator of the system induced by the Hamiltonian $\hat{H}_S+\hat{H}_I$. Notice that $u$ has to be chosen as a real number (and in particular proportional to a time $t$) to attain $\mathcal{G}(u,\Delta o)$ using an interferometric scheme, as we will discuss in the next section.

\section{Loschmidt echo and quasiprobability distribution of work}\label{sec:LE_Pw}

Let us consider the quench of a fermionic Hamiltonian from $\hat{H}_S$ to $\hat{H}_S^{'}=\hat{H}_S + \hat{H_I}$, and that the two operators admit the following spectral decompositions:
\begin{eqnarray}
    \hat{H}_S &=& \sum_{n}E_{n}\ketbra{\Psi_n}{\Psi_n} \\
    \hat{H}_S^{'} &=& 
    \sum_{m}E^{'}_{m}\ketbra{\Psi_m^{'}}{\Psi_m^{'}}.
\end{eqnarray}
The Loschmidt echo $\nu(t) \equiv {\rm Tr}[\hat{U}'(t) \hat{\rho} \hat{U}(-t)]$, applied to a generic density operator $\hat{\rho}$ with $\hat{U}(t) = e^{-i\hat{H}_S t}$ and $\hat{U}'(t) = e^{-i\hat{H}_S^{'} t}$, is the characteristic function of the KDQs $q_{n,m}$ of the random variable $w_{n,m} \equiv E'_m-E_n$~\cite{GherardiniTutorial}. Formally, 
\begin{equation}\label{eq:LE_char_func}
    \nu(t) = {\rm Tr}\left[ \hat{U}'(t) \hat{\rho} \hat{U}(-t) \right] = \sum_{n,m} q_{n,m} e^{-i(E'_m-E_n)t}. 
\end{equation}
Notice that $\hat{U}'$ and $\hat{U}$ are the unitary operators that describe the time-evolution of the fermionic system perturbed and not perturbed by the defect, respectively. We assume that the system is initialized in a pure state, expressed, without loss of generality, with respect to the eigenbasis of $\hat{H}_S$. This means that
\begin{equation}\label{eq:rho_decomposition}
    \hat{\rho} = \ketbra{\Theta}{\Theta} = \sum_{k,\ell} \alpha_k \alpha^*_{\ell} \ketbra{\Psi_k}{\Psi_{\ell}},
\end{equation}
where $\ket{\Theta} = \sum_{j}\alpha_{j}\ket{\Psi_j}$ with the coefficients $\alpha_j$ fulfilling the normalization condition $\sum_j|\alpha_j|^2=1$. If we insert the explicit expression of $\hat{\rho}$ defined in \eqref{eq:rho_decomposition} in the LE $\nu(t)$, it becomes
\begin{eqnarray}\label{eq:LE}
    \nu(t) = \sum_{n,m,k} e^{-i (E'_m - E_n) t} \alpha_k \alpha^*_n \Lambda^*_{m,n} \Lambda_{m,k}\,,
\end{eqnarray}
where we have used the eigenbasis $\ket{\Psi}$ and $\ket{\Psi'}$ of $\hat{H}_S$ and $\hat{H}^{'}_S$ respectively, and we have defined the overlaps $\Lambda_{m,n} \equiv \braket{\Psi'_m}{\Psi_n}$. Interestingly, if we compare \eqref{eq:LE} with the definition of $\nu(t)$ in \eqref{eq:LE_char_func}, we obtain an explicit expression of the KDQ $q_{n,m}$ as a function of the overlaps $\Lambda_{m,n}$:
\begin{equation}\label{eq:q_mn}
    q_{n,m} = \alpha_n^* \Lambda^*_{m,n} \sum_{k} \Lambda_{m,k} \alpha_k \,.
\end{equation}
We thus define the KDQ distribution
\begin{equation}\label{eq:work_distribution}
    P[w] = \sum_{n,m} q_{n,m} \delta(w - w_{n,m})
\end{equation}
of the work done by the defect on the fermionic system. The non-positivity functional quantifies the extent of how much each KDQ $q_{n,m}$ can be real negative or even a complex number, and is defined as $\mathcal{N}\equiv -1 + \sum_{n,m} \abs{q_{n,m}}$~\cite{GherardiniTutorial}. 
The non-positivity of the work distribution $P[w]$ comes from the presence of quantum coherence in the initial state $\hat{\rho}$ in the eigenbasis of $\hat{H}_S$. Quite different results are obtained if the initial state of the system is diagonal in $\hat{H}_S$:  
\begin{equation}
    \hat{\rho} = \sum_{k} \beta_k \ketbra{\Psi_k}{\Psi_k}, 
\end{equation}
where $\beta_k \equiv |\alpha_k|^2$. Using this initial state, the Loschmidt echo $\nu(t)$ is
\begin{eqnarray}
    \nu(t) = \sum_{n,m} e^{-i (E'_m - E_n) t} \beta_n |\Lambda_{m,n}|^2
\end{eqnarray}
with the result that KDQ $q_{n,m}$ are real and positive numbers belonging to the interval $[0,1]$, as in \cite{goold_orthogonality_2011}. In particular, in such a case, the $q_{n,m}$ assume the simple form
\begin{equation}\label{q_mn_diagonal}
    q_{n,m} = \beta_n |\Lambda_{m,n}|^2 = |\alpha_n|^2  |\Lambda_{m,n}|^2 
\end{equation}
that can be determined using the (standard) two-point measurement scheme~\cite{CampisiRMP2011}. Having $q_{n,m}$ probabilities respecting the theory of classical probability, the non-positivity functional is identically equal to zero.

\subsection{Ramsey interferometric scheme for measuring Loschmidt echo}\label{sec:Ramsey_scheme}

The effects of the presence of the localized defect can be measured using quantum interferometry. We recall that we are assuming that the fermionic system is perturbed by the defect only when the latter is in a specific state, for example $\hat{H}=\hat{H}_0+\ketbra{1}{1} \otimes \hat{V}$, as discussed in \cite{goold_orthogonality_2011}.

\begin{figure}[b]
\centering
    \begin{quantikz}[slice all]
        \ket{0} & \gate{H} & \ctrl{1} & \gate{\hat{R}_{\varphi}} & \gate{H} & \meter{\sigma_z} \\
        \ket{\Psi_g} & & \gate{\hat{U}_c} & & & 
    \end{quantikz}
    \caption{Circuit representation of the Ramsey interferometric scheme for measuring the real and imaginary parts of the Loschmidt echo $\nu(t)$. Both the defect and the system are initially taken in their ground states. Here, $H$ and $\hat{R}_{\varphi}$ denote the Hadamard and quantum phase gate respectively, while $\hat{U}_c$ is the controlled-unitary gate of \eqref{eq:controlled-unitary}.}
    \label{fig:RamseyScheme}
\end{figure}
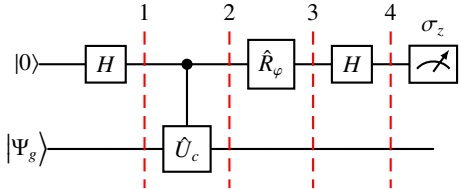

Let us start by considering both the defect and the fermionic system in their ground states, such that the state of the composite system is $\ket{\Xi_0} = \ket{0} \otimes \ket{\Psi_g}$. At time $t_1$, a Hadamard gate (identified by $H$ in Fig.~\ref{fig:RamseyScheme}) is applied to the defect (e.g., by turning on the interaction with a laser), so that $\ket{\Xi(t=t_1)} = \left(\ket{0}+\ket{1}\right)/\sqrt{2} \otimes \ket{\Psi_g}$. Then, the system evolves for a time $\tau$ under the effect of the interaction Hamiltonian $\hat{H}_I$ according to the following controlled-unitary quantum gate:
\begin{equation}\label{eq:controlled-unitary}
    \hat{U}_c = \ketbra{1}{1} \otimes e^{-i(\hat{H}_S + \hat{H_I})\tau} + \ketbra{0}{0} \otimes e^{-i\hat{H}_S\tau}.
\end{equation}
The resulting state of the composite system at $t=t_2=t_1+\tau$ is (Eq.~(7) in \cite{goold_orthogonality_2011}):
\begin{equation}
    \ket{\Xi(t=t_2)} =
    \frac{1}{\sqrt{2}} \left( \ket{0} \otimes e^{-i\hat{H}_S\tau} \ket{\Psi_g} + \ket{1} \otimes e^{-i(\hat{H}_S + \hat{H_I}) \tau} \ket{\Psi_g} \right).
\end{equation}
After the controlled-unitary gate, the reduced density operator of the defect (obtained by tracing out the system) is 
\begin{equation}
\hat{\rho}_D(t_2) = \frac{1}{2} 
\begin{pmatrix}
   1 & \nu^*(t_2) \\
   \nu(t_2) & 1 
\end{pmatrix},
\end{equation}
where $\nu(t) = \bra{\Psi_g}e^{i\hat{H}_S t}e^{-i(\hat{H}_S + \hat{H_I})t}\ket{\Psi_g}$ is the Loschmidt echo.

As we aim to measure the real and imaginary parts of $\nu(t)$, we introduce a further tunable parameter, namely the phase $\varphi$ that is applied to the defect at $t=t_2$ via the phase gate $\hat{R}_{\varphi} = \begin{pmatrix} 1 & 0 \\ 0 & e^{i\varphi} \end{pmatrix}$. Then, at $t=t_3$, we perform a second Hadamard gate that closes the interferometer. This means that, if the phase $\varphi=0$ and there is no interaction ($k=0$) between the fermionic system and the defect, then the state at the end of the interferometer at $t=t_4$ is the same of that at the beginning. The complete expression for the final state of the composite system reads as
\begin{eqnarray}
    \ket{\Xi(t=\tau)} = \frac{1}{\sqrt{2}} \Bigg[ \left(\frac{\ket{0}_D+\ket{1}_D}{\sqrt{2}} \right) \otimes e^{-i\hat{H}_S\tau} \ket{\Psi_g} \notag \\
    + e^{i\varphi} \left(\frac{\ket{0}_D-\ket{1}_D}{\sqrt{2}} \right) \otimes e^{-i(\hat{H}_S + \hat{H_I})\tau} \ket{\Psi_g} \Bigg]\,.    
\end{eqnarray}
Measuring in the eigenbasis of the observable $\hat{\sigma}_z$, we determine the probability of finding the defect in its ground state:
\begin{eqnarray}
    p_g(\tau,\varphi) &=& \frac{ \left(1 + \text{Re}\left\{ e^{i\varphi}\nu(\tau)\right\} \right) }{2} = \notag \\
    &=& \frac{ \left( 1 + \cos(\varphi) \text{Re}\left\{ \nu(\tau) \right\} -\sin(\varphi) \text{Im}\left\{ \nu(\tau) \right\} \right) }{2} \,.    
\end{eqnarray}
This means that, by properly changing the value of $\varphi$, we can reconstruct the real and imaginary parts of the Loschmidt echo $\nu(\tau)$ as a fitting parameter from measurements of $\hat{\sigma}_z$. 

In the next section, we will show how we can determine the overlaps $\Lambda$ between the many-body wave-functions decomposing the free-fermions Hamiltonian.  

\section{Overlap of many-body fermionic wave-functions}
\label{sec:overlap_many-body_state}

The many-body wave-function, projected in the 1D space of positions $x$, of $N$ free-fermions is given by using the Slater determinant:
\begin{equation}\label{eq:proj_many-body_Slater}
    \Psi(X) = \frac{1}{\sqrt{N!}} \sum_P \text{sgn}(P) \ \psi_{P(1)}(x_1) \psi_{P(2)}(x_2) \dots \psi_{P(N)}(x_N),
\end{equation}
where the sum is extended over all the possible permutations $P$ of the indices labelling the occupied states by each fermion in one of the $N$ positions $X = (x_1,x_2,\ldots,x_N)$. The overlap between the many-body wave-function without and with (denoted by an apostrophe as superscript) the presence of the defect assumes the following expression:
\begin{align}
    \Lambda_{m,n} &= \braket{\Psi_m'(X)}{\Psi_n(X)} = \int \left[ \Psi_m'(X) \right]^{*} \Psi_n(X) \, dX = \nonumber \\
    &= \sum_{P,P'} \text{sgn}(P) \text{sgn}(P') \prod_{\ell} A_{P(\ell),P'(\ell)}\,,
\end{align}
where 
\begin{equation}\label{eq:def_Aij}
A_{m,n} \equiv \int [\psi_m'(x)]^* \psi_n(x) \, dx \,.
\end{equation}
The elements of the matrix $A$ are the overlaps between the eigenstates of the single-particle Hamiltonians.

In order to calculate the overlap between the many-body wave-functions, we introduce the operator $\Hat{P}_{-}$ that anti-symmetrize the wave-function it is applied to, as requested by the computation of the Slater determinant. In particular,
\begin{equation}\label{P_def}
    \hat{P}_{-} \equiv \frac{1}{N!} \sum_P {\rm sgn}(P) \hat{W}_P \,,
\end{equation}
where $P$ denotes all the possible permutations and the operator $\hat{W}_P$ permutes the indices of the fermionic many-body wave-function as $\hat{W}_P(\psi_1, \ldots,\psi_N) = \psi_{P(1)} \psi_{P(2)} \cdots\psi_{P(N)}$. 
The operator $\Hat{P}_{-}$ is Hermitian and is a projector, as demonstrated in \cite{Bratteli1997}. Thus, using the notation
\begin{equation}\label{eq:Psi_Slater}
    \ket{\hat{P}_- \Psi} = \frac{1}{\sqrt{N!}} \ket{\Psi_{\rm Slater}} 
\end{equation}
where $\ket{\Psi_{\rm Slater}}$ denotes the many-body wave-function vector whose projection on the position space is $\Psi(X)$ of \eqref{eq:proj_many-body_Slater}, the overlap between two generic $\ket{\Psi_{\rm Slater}}$ before and after the quench transformation simplifies to 
\begin{eqnarray}\label{eq:Simplification_Lambda}
    \Lambda &=& \bra{\Psi'_{\rm Slater}} \ket{\Psi_{\rm Slater}} = N! \bra{\Psi'\hat{P}_-}\ket{\hat{P}_- \Psi} = N! \bra{\Psi'} \ket{ \hat{P}_-\hat{P}_- \Psi} = \notag \\
    &=& N! \bra{\Psi'} \ket{ \hat{P}_- \Psi} = \sqrt{N!} \bra{\Psi'} \ket{\Psi_{\rm Slater}}.
\end{eqnarray}
In the second step of the calculation for \eqref{eq:Simplification_Lambda} we have used the fact that $\hat{P}_-$ is Hermitian, while in the third step the property that it is a projector. Using \eqref{eq:Simplification_Lambda} reduces the complexity, in terms of the number of operations, for computing the overlap $\Lambda$. Indeed, from \eqref{eq:Simplification_Lambda}, we evince that one has to perform only the anti-symmetrization of the unperturbed wave-function vector. Hence, adopting the simplification in \eqref{eq:Simplification_Lambda}, the matrix $\Lambda$ containing the single-particle overlaps $\Lambda_{m,n}$ explicitly reads as
\begin{eqnarray}\label{Lambda_Slater_determinant_matrix}
    \Lambda &=& \sqrt{N!} \bra{\Psi'} \ket{\Psi_{\rm Slater}} = \notag \\
    &=& \sqrt{N!} \int dX \, \Psi'^*(X) \frac{1}{\sqrt{N!}} \sum_{P } \text{sgn}(P) \prod_{i=1}^N \psi_{P(i)}(x_i) = \notag \\
    &=& \sum_{P } \text{sgn}(P) \prod_{i=1}^N \int [\psi_i'(x_i)]^* \psi_{P(i)}(x_i) \, dx_i = \notag \\
    &=& \sum_{P } \text{sgn}(P) \prod_{i=1}^N A_{i,P(i)} = \det(A) \,.
\end{eqnarray}
As a remark, note that the single-particle wave-functions $\psi_1,\ldots,\psi_N$ do not necessarily need to be the first $N$ eigenstates of the Hamiltonian they refer to; indeed, they can be any list of $N$ eigenstates. 

\subsection{Analytical expression of the overlaps in the limit of strong perturbation}\label{eq:overlaps_strong_k}

Using \eqref{Lambda_Slater_determinant_matrix}, the overlap between many-body wave-functions can be determined starting from the overlaps of the single-particle energy eigenstates. Here, we show that these can be computed analytically. In the limit of strong interactions $k \rightarrow +\infty$, the analytical expression of the
perturbed and unperturbed Hamiltonians' eigenstates are
\begin{eqnarray}
&&\psi_n(x) = \frac{1}{(2\pi)^{1/4}} \frac{1}{\sqrt{2^n n!}} e^{-x^2/4} H_n \left( \frac{x}{\sqrt{2}} \right),\label{eq:unperturbed_psi}\\
&&\psi'_n(x) = \nonumber\\
&&\begin{cases}
\displaystyle{ \psi_{n+1}(\abs{x}) = \frac{1}{(2\pi)^{1/4}} \frac{1}{\sqrt{2^{n+1} (n+1)!}} e^{-\abs{x}^2/4} H_{n+1} \left( \frac{\abs{x}}{\sqrt{2}} \right),} & \text{n even} \\
\displaystyle{ \psi_{n}(x) = \frac{1}{(2\pi)^{1/4}} \frac{1}{\sqrt{2^n n!}} e^{-x^2/4} H_n \left( \frac{x}{\sqrt{2}} \right),} & \text{n odd} ,
\end{cases}\nonumber\\
&&\label{eq:perturbed_psi}
\end{eqnarray}
where $H_n$ denotes the Hermite polynomial of order $n$. In the calculations, it is worth considering the following symmetries:
\begin{equation}\label{eq:Lambda_mn_cases}
    \Lambda_{m,n} = \braket{\psi_m'}{\psi_n} = 
    \begin{cases}
        \braket{\psi_m}{\psi_n} = \delta_{m,n}, & \text{m odd} \\
        \braket{\psi_m'}{\psi_n} = 0, & \text{m even, n odd} \\
        \braket{\psi_m'}{\psi_n} , & \text{m even, n even} \,.
    \end{cases}
\end{equation}
Hence, one has to consider only the even energy eigenstates. In conclusion, the general expression for the overlaps in the strong perturbation limit assumes the form (see also Appendix \ref{app:Appendix_strong_pert} for details)
\begin{equation}\label{eq:Lambdamn_strong}
    \Lambda_{m,n} = (-1)^{n/2} \Lambda_{m,0} \sqrt{\frac{(n-1)!!}{n!!}} \frac{m+1}{m+1-n} \,,
\end{equation}
where
\begin{equation}\label{eq:LambdaM0strong}
    \Lambda_{m,0} = (-1)^{m/2} \sqrt{\frac{2^{m+1}}{(m+1)!}} \frac{\Gamma \left(\frac{m+1}{2} \right)}{\pi} \,.
\end{equation}
Finally, a more tractable expression is given by the recursive relation
\begin{equation}\label{eq:LambdaMNstrong_recursive}
    \frac{\Lambda_{m,n}}{\Lambda_{m,n-2}} = - \sqrt{\frac{n-1}{n}} \frac{m-n+3}{m-n+1} \,.
\end{equation}

\begin{figure}[t]
    \centering
    \includegraphics[width=\columnwidth]{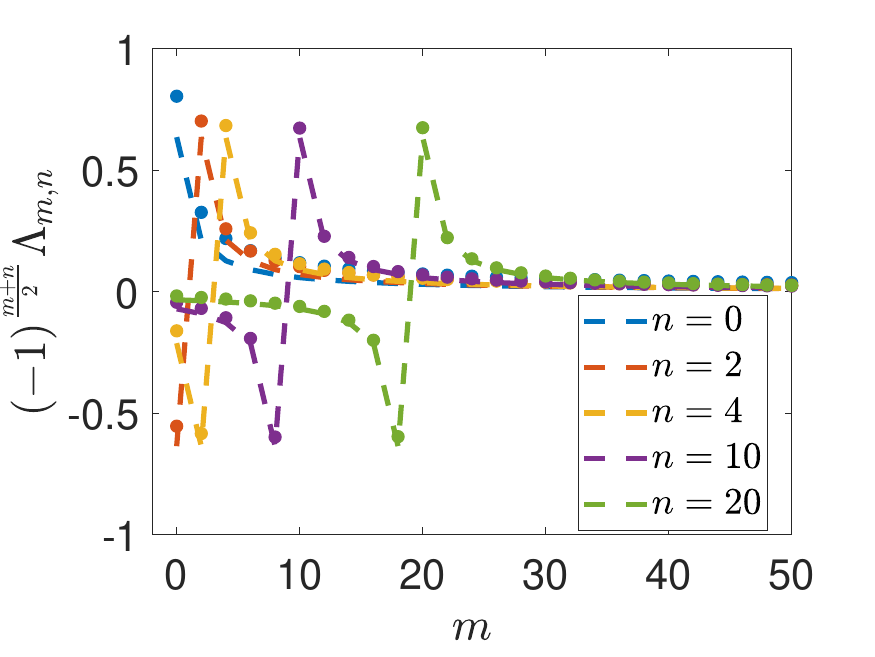}
    \caption{Comparison of the overlaps $\Lambda_{m,n}$, computed numerically for $k=100$, with the theoretical prediction of \eqref{eq:Lambdamn_hyperbola} that is valid for large $n$. Both curves are plotted as a function of $m$, with $n=0,2,4,10,20$. 
    For clarity of illustration, the overlaps are multiplied by the factor $(-1)^\frac{m+n}{2}$ in order to remove the alternating sign. Dots represent the numerical values of the overlaps, while dashed lines are the corresponding approximations for large $n$.}
    \label{fig:Lambda_m0_m2_m20}
\end{figure}

\subsubsection*{Approximation for large $n$}

The expression of the overlap $\Lambda_{m,n}$ in the limit of strong interaction $k \rightarrow +\infty$ can be approximated using the Stirling formula for the factorial: $n! \approx \sqrt{2\pi n}\,(n/e)^n$, which is valid for large $n$. In this way, the overlap resembles the discretised version of a hyperbola function with alternating signs, i.e.,
\begin{equation}\label{eq:Lambdamn_hyperbola}
    \Lambda_{m,n} \approx \frac{ (-1)^{\frac{m+n}{2}} }{\pi} \frac{2}{m-n+1}\,.
\end{equation}
In Fig.~\ref{fig:Lambda_m0_m2_m20} we compare the right-hand-side of \eqref{eq:Lambdamn_hyperbola} (dashed lines) with the overlaps $\Lambda_{m,n}$ computed numerically for $k=100$ (dots connected by dotted lines), as a function of $m$, with $n=0,2,4,10,20$. We can observe that, except for $n=0$, the Stirling approximation well reproduces the numerical overlaps. Moreover, in accordance with \eqref{eq:Lambda_mn_cases}, all overlaps are zero for odd values of $m$ and $n$, while for increasing $n$ the hyperbola of \eqref{eq:Lambdamn_hyperbola} shifts proportionally to the value of $m$. It is worth noting that \eqref{eq:Lambdamn_hyperbola} cannot be properly considered as a hyperbola, as it represents a discrete function of the variables $n$ and $M$. However, the asymptotic discontinuity of the function is still manifest in the fact that the overlap $\Lambda_{m,n}$ diverges for $m = n-1$. As we will see later on, the composition of many such overlaps, each exhibiting singular behaviour, can give rise to discontinuities in the LE, which resembles trends discussed in Ref.~\cite{Knap2012} for a genuinely many-body context. Moreover, notice that for systems composed of a few fermions initialized in a superposition of the ground state and at least one excited state of the unperturbed Hamiltonian, the overlaps $\Lambda_{m,n}$ contribute differently depending on which excited states are included in the initial quantum superposition.

\begin{figure}[t]
    \centering
    \includegraphics[width=\columnwidth]{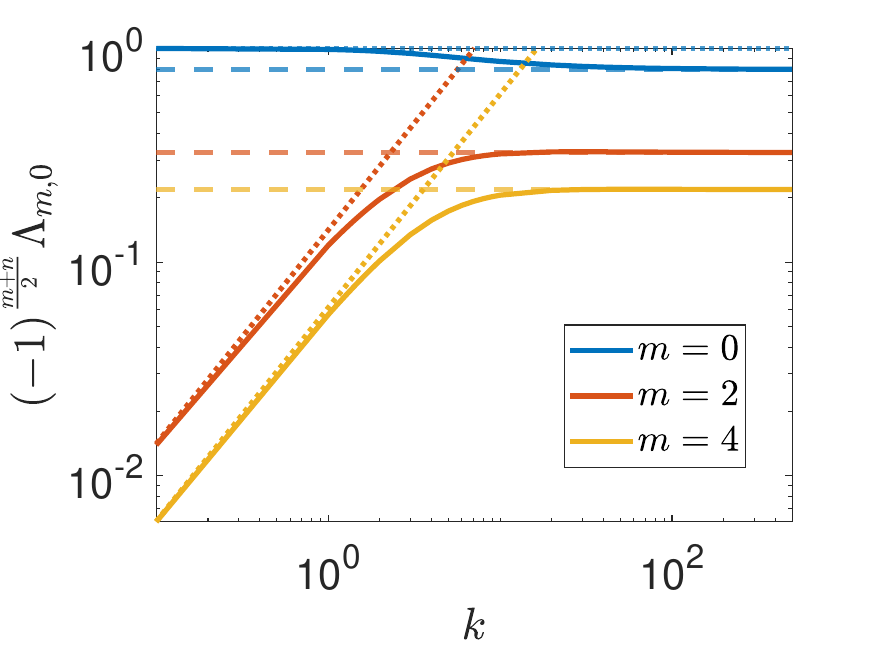}
    \caption{Plot of the overlaps $\Lambda_{0,0}$, $\Lambda_{2,0}$ and $\Lambda_{4,0}$ as a function of $k$ (solid lines), obtained from direct numerical computations. As in Fig.~\ref{fig:Lambda_m0_m2_m20}, the overlap is multiplied by the factor $(-1)^\frac{m+n}{2}$ in order to remove the alternating sign. Dotted lines represent the approximation for $k \rightarrow 0$, while the dashed lines denote the limit value for $k$ large.}
    \label{fig:overlap_small_k}
\end{figure}

\subsection{Analytical expression of the overlaps in the limit of weak perturbation}

In the weak perturbation limit $k \rightarrow 0$, the eigenfunctions and eigenvalues of the perturbed Hamiltonian $\hat{H}_S + \hat{H}_I$ can be still computed analytically using first-order perturbation theory: 
\begin{align}
    \ket{\Psi_m'^{(0)}} &= \ket{\Psi_m} + \sum_{i \neq m} \frac{\bra{\Psi_i} \hat{H_I} \ket{\Psi_m}}{E_m - E_i} \ket{\Psi_i} = \nonumber \\
    &= \ket{\Psi_m} + \sum_{i\neq m} \frac{1}{m-i} \frac{k}{\sqrt{2\pi}} c_m c_i \ket{\Psi_i},\label{eq:psi_E_perturbed_weak_1} \\    
    E_m'^{(0)} &= E_m + \bra{\Psi_m} \hat{H_I} \ket{\Psi_m} = m+ \frac{1}{2} + \frac{k}{\sqrt{2\pi}} c_m^2 \,,\label{eq:psi_E_perturbed_weak_2}
\end{align}
where $c_n=(n-1)!!/\sqrt{n!}$, and the superscript $(0)$ denotes the limit $k \rightarrow 0$. Thus, using Eqs.~(\ref{eq:psi_E_perturbed_weak_1})-(\ref{eq:psi_E_perturbed_weak_2}), the overlaps and energy differences simplify to
\begin{align}\label{eq:overlap_energydiff_weak}
    \Lambda_{m,n}^{(0)} &= \braket{\Psi_m'^{(0)}}{\Psi_n} = 
    \begin{cases}
        1, & m = n, \\
        \displaystyle{\frac{k}{m-n} \frac{1}{\sqrt{2\pi}} c_n c_m}, & m \neq n,
    \end{cases} \\
    \Delta E_{m,n}^{(0)} &= E_m'^{(0)} - E_n = m - n + \frac{k}{\sqrt{2\pi}} c_m^2.
\end{align}
This perturbative approach successfully applies to the calculation of expectation values, like the overlaps, as well as of quasiprobabilities that are obtained by substituting the expression of $\Lambda_{m,n}^{(0)}$ given by \eqref{eq:overlap_energydiff_weak} into $q_{n,m} = \alpha_n^* \Lambda_{m,n}^{*(0)} \sum_{j} \Lambda_{m,j}^{(0)} \alpha_j$. In the regime of weak perturbations, it can be determined that $\sum_{n,m} q_{n,m} = 1 + O(k^2)$ as expected.

We conclude by plotting in Fig.~\ref{fig:overlap_small_k} the overlap $\Lambda_{m,0}$ ($m=0,2,4$) as a function of the perturbation strength $k$ (solid lines), obtained numerically as in Fig.~\ref{fig:Lambda_m0_m2_m20}. These numerical results agree with both the analytical expressions of \eqref{eq:overlap_energydiff_weak} for $k\rightarrow 0$ (dotted lines) and of \eqref{eq:Lambdamn_strong} for $k\rightarrow +\infty$ (dashed lines).  

\begin{figure*}[t]
    \centering
    \includegraphics[width=0.66\columnwidth]{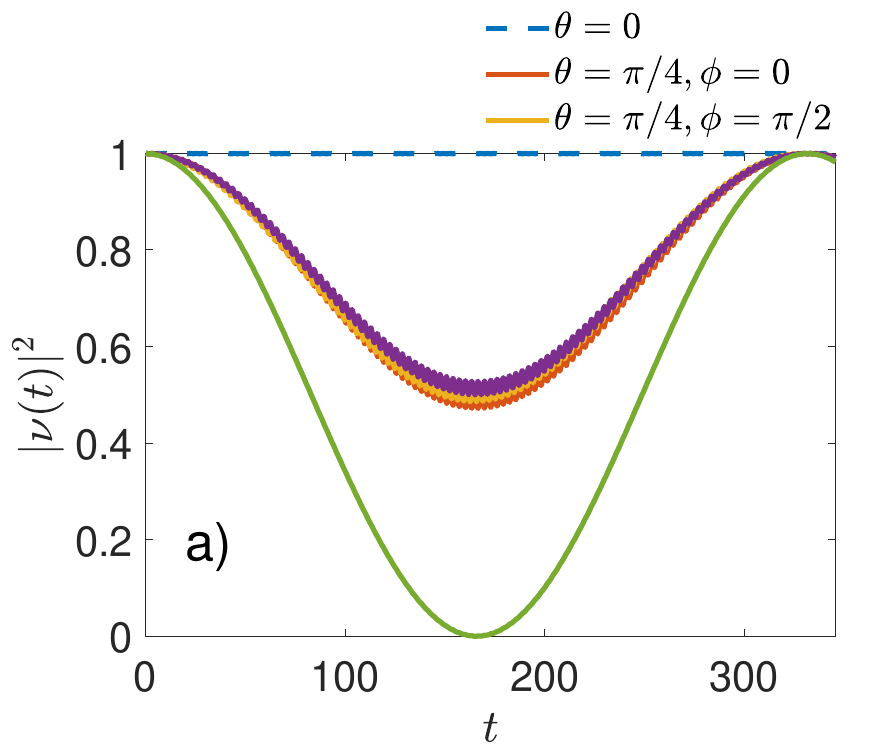}
    \includegraphics[width=0.66\columnwidth]{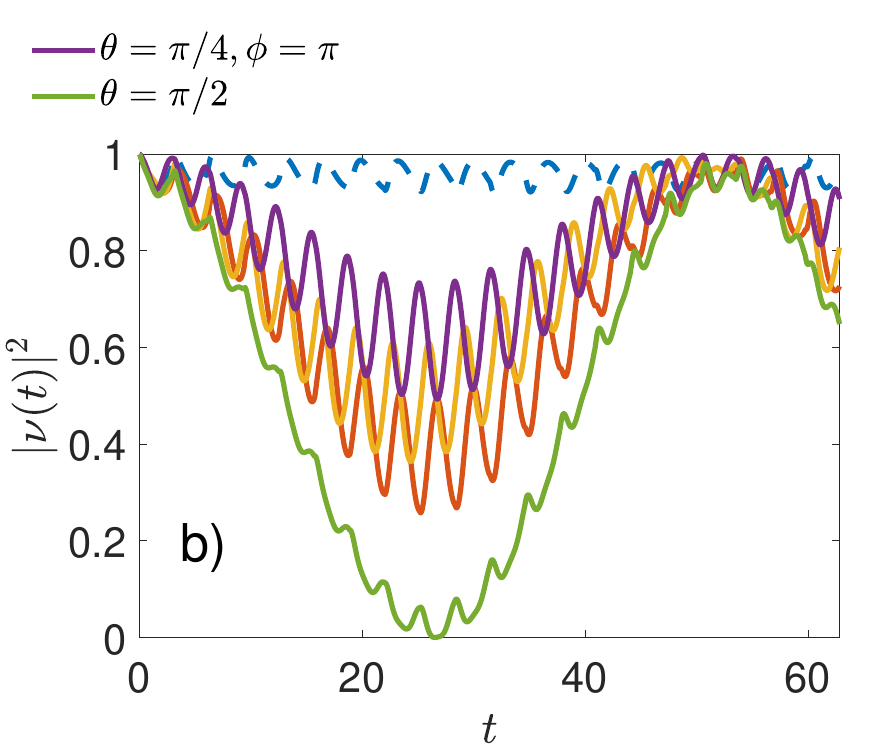}
    \includegraphics[width=0.66\columnwidth]{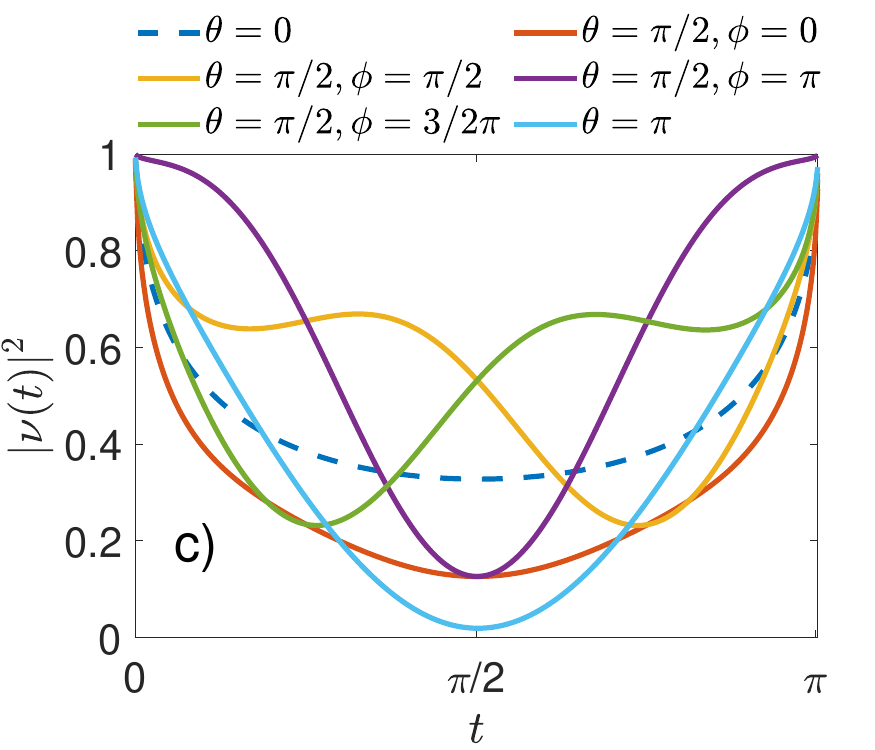}
    \caption{Loschmidt echo $|\nu(t)|^2$ as a function of time $t$ for initial states with $n_1=0$, $n_2=2$ and different angles $\theta$ and $\phi$, as defined in \eqref{eq:Bloch_state}. From left to right the interaction strength is $k=0.1,1,100$.
    }
    \label{fig:LoschmidtEcho_BlochState_various_k}
\end{figure*}

\section{Average and variance of work done by the defect}\label{sec:Average_variance_work}

Let us consider that the fermionic system is initialized in the state corresponding to the density operator $\hat{\rho}=\ketbra{\Psi_0}{\Psi_0}$ containing quantum coherence with respect to the unperturbed Hamiltonian $\hat{H}_S$ before interacting with the localized defect. In particular, we take $\ket{\Psi_0}$ as a combination of $N$ eigenstates of $\hat{H}_S$. Moreover, for brevity, from here on, we will denote the Hamiltonian operators $\hat{H}_S, \hat{H}_S + \hat{H_I}$ respectively as $\hat{H}_i, \hat{H}_f$ with $i, f$ standing for the initial and final times of the quench transformation $\hat{H}_S \rightarrow \hat{H}_S + \hat{H_I}$.

We thus compute the average work associated to the quench transformation:
\begin{align}
     \langle w \rangle &= 
     {\rm Tr}\left[ ( \hat{H}_f - \hat{H}_i ) \, \hat{\rho} \right] = 
     \sum_{n,j=1}^N \alpha_n^* \alpha_j \bra{\Psi_n} ( \hat{H}_f - \hat{H}_i ) \ket{\Psi_j} = \nonumber \\
     &= \sum_{n,j=1}^N \alpha_n^* \alpha_j \bra{\Psi_n} ( \hat{H}_f - \hat{H}_i ) \left( \sum_{m=0}^{\infty} \ketbra{\Psi_m'}{\Psi_m'} \right) \ket{\Psi_j} = \nonumber \\
     &= \sum_{n,j=1}^N \sum_{m=0}^{\infty} \alpha_n^* \alpha_j (E_m'- E_n) \braket{\Psi_n}{\Psi_m'} \braket{\Psi_m'}{\Psi_j} =  \nonumber \\
     &= \sum_{n=1}^N \sum_{m=0}^{\infty} (E_m' - E_n) \alpha_n^* \Lambda_{m,n}^* \sum_{j=1}^N \alpha_j \Lambda_{m,j} = \sum_{n,m} \Delta E_{n, m} \ q_{n,m}.
\end{align}
For clarity, we recall that $\ket{\Psi_n}$ and $\ket{\Psi_m'}$ are the many-body eigenstates of the unperturbed and perturbed Hamiltonian operators, $\hat{H}_i$ and $\hat{H}_f$ respectively.

Concerning the second statistical moment of the work distribution, we observe that the formula (valid also in the non-commutative case)~\cite{Fusco2014}
\begin{equation}\label{eq:W_second_moment}
   \langle w^2 \rangle = \bra{\Psi_0} \left( \hat{H}_f^2 - 2 \hat{H}_f \hat{H}_i + \hat{H}_i^2 \right) \ket{\Psi_0}  
\end{equation}
gives the same result of the variance of work associated to the KDQ distribution $P[w]$, which assumes the following expression:
\begin{equation}
    \langle w^2 \rangle = \sum_{n,m} \left( E_m'^2 - 2E_m'E_n + E_n^2 \right) q_{n,m} \,.
\end{equation}
For an initial state given by the superposition of unperturbed eigenstates [see \eqref{eq:rho_decomposition}], the second statistical moment of work is linked with the average work through the relation
\begin{equation}\label{eq:second_moment_work}
    \langle w^2 \rangle = \frac{k}{\sqrt{2 \pi}} \left( 1 + \frac{\zeta(1/2)}{\sqrt{\pi}} \right) \langle w \rangle \,,
\end{equation}
where $\zeta(s)$ is the Riemann zeta function. Albeit the analytical continuation of $\zeta(s)$ gives a finite number for $0<s<1$, the summation in the definition of the Riemann zeta function diverges for $s<1$. As a result, a divergent behaviour is expected for $\langle w^2 \rangle$ when initializing the system in every pure superposition state.

Below, we are going to present two paradigmatic case-studies, for the single particle and two fermions, that clearly illustrate --- with analytical and numerical results --- the effects entailed by a delta-perturbation in the general case the system is prepared in a superposition of energy levels of the Hamiltonian before the quench transformation.

\section{Case-study 1: Single particle}\label{sec:single_particle}

\subsection{Analysis with a qubit}

We begin our case-study analysis from the simplest scenario, namely a single particle ($N=1$). In particular, we start considering as initial state $\ket{\Psi_0}$ a coherent superposition of $d=2$ eigenstates of the unperturbed Hamiltonian $\hat{H}_i$, i.e., a qubit. In order to calculate the KDQ of the work distribution, we need to take into account, in principle, all the eigenstates of the perturbed Hamiltonian $\hat{H}_f$. This means that the sum over the index $m$ in \eqref{eq:q_mn} has to run over all the eigenstates of $\hat{H}_f$. Albeit simple, this choice ensures that $\hat{\rho}=\ketbra{\Psi_0}{\Psi_0}$ does not commute with $\hat{H}_i$ and likely with $\hat{H}_f$ either, thus capturing non-trivial interference effects due to quantum coherence.

We choose
\begin{equation}\label{eq:ket_Psi0}
    \ket{\Psi_0} = \alpha_1 \ket{\Psi_{n_1}^{(i)}} + \alpha_2 \ket{\Psi_{n_2}^{(i)}},
\end{equation}
where $\alpha_1,\alpha_2$ are complex coefficients and $\ket{\Psi_{n_1}^{(i)}},\ket{\Psi_{n_2}^{(i)}}$ are two eigenstates of $\hat{H}_i$. The associated KDQ $q_{n,m}$ of \eqref{eq:q_mn} reduce to the following expressions:
\begin{align}
    q_{n_1,m} &= \abs{\alpha_1}^2 \Lambda_{m,n_1}^2 + \alpha_1^* \alpha_2 \Lambda_{m,n_1} \Lambda_{m,n_2} \label{eq:q_1m} \\
    q_{n_2,m} &= \alpha_1 \alpha_2^* \Lambda_{m,n_1} \Lambda_{m,n_2} + \abs{\alpha_2}^2 \Lambda_{m,n_2}^{2}, \label{eq:q_2m}
\end{align}
where we have used the reality of the overlaps $\Lambda_{m,n}$. Eqs.~\ref{eq:q_1m}-\ref{eq:q_2m} represent the probabilities that the quantum system ``jumps'' from the eigenstates $\ket{\Psi_{n_1}^{(i)}}$ or $\ket{\Psi_{n_2}^{(i)}}$ of the unperturbed Hamiltonian $\hat{H}_i$ to an eigenstate $\ket{\Psi_m'}$ of $\hat{H}_f$. In the limit where the initial state is exactly an eigenstate of $\hat{H}_i$, i.e.~$\alpha_1=0$ or $\alpha_2=0$, the KDQ reduce to the classical transition probabilities $p_{n_1,m} = \abs{\alpha_1}^2 \Lambda_{m,n_1}^2$ and $p_{n_2,m} = \abs{\alpha_2}^2 \Lambda_{m,n_2}^2$
as determined by the two-point measurement scheme. Similarly, in the absence of perturbation ($k = 0$), the overlaps $\Lambda_{m,n}$ become Kronecker deltas ($\Lambda_{m,n} = \delta_{m,n}$), and the KDQ assume the constant values $q_{n_1,m} = |\alpha_1|^2$ and $q_{n_2,m} = |\alpha_2|^2$.

The key quantities in Eqs.~(\ref{eq:q_1m})-(\ref{eq:q_2m}) are the overlaps $\Lambda_{m,n_1}$ and $\Lambda_{m,n_2}$. Since the odd eigenstates of $\hat{H}_i$ are not affected by the perturbation, their overlaps with the eigenstates of $\hat{H}_f$ are non-zero only for $m=n$. Therefore, from here on, we restrict our attention to the superposition of even eigenstates of $\hat{H}_i$. To frame our analysis in more geometric terms, we parametrize the two-level state $\ket{\Psi_0}$ on a Bloch sphere as
\begin{equation}\label{eq:Bloch_state}
    \ket{\Psi_0} = \cos\left(\frac{\theta}{2}\right) \ket{n_1} + e^{i\phi} \sin\left(\frac{\theta}{2}\right) \ket{n_2},
\end{equation}
where $\theta \in [0,\pi]$ and $\phi \in [0,2\pi)$ are the polar angles in the Bloch sphere. In the specific case studied here, we set $\ket{n_1} = \ket{0}$ as the ground state and $\ket{n_2} = \ket{2}$ as the first even excited state of $\hat{H}_i$. By fixing the perturbation strength to $k = 100$ and initializing the particle in a superposition of the ground state of $\hat{H}_i$ and its second excited state (the first even eigenstate), the overlap functions $\Lambda_{m,0}$ and $\Lambda_{m,2}$ obtained numerically are shown in Fig.~\ref{fig:Lambda_m0_m2_m20}. Due to the symmetry of the defect, overlaps with odd eigenstates are always equal to zero. 

\begin{figure*}[ht]
    \centering
    \includegraphics[width=0.33\linewidth]{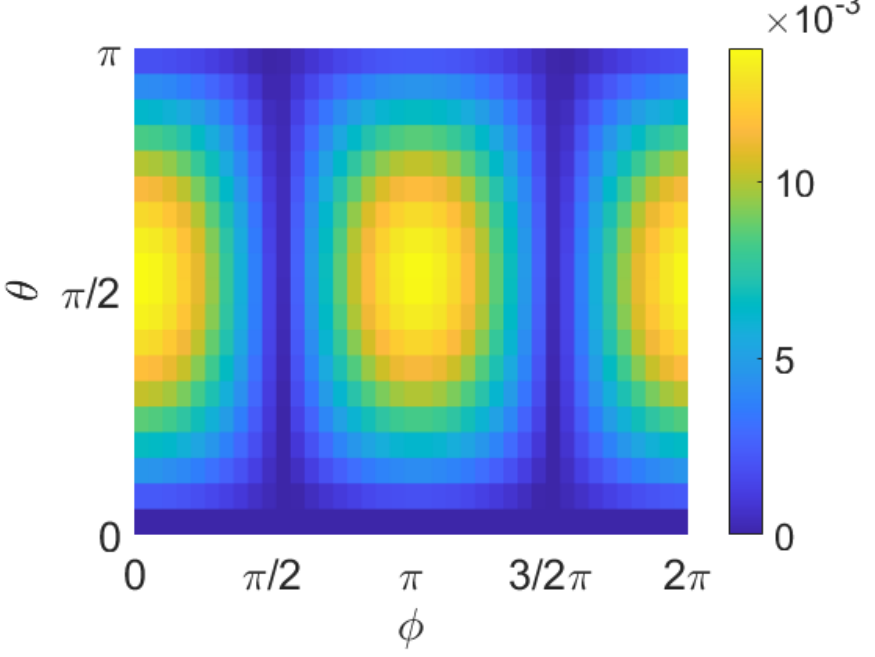}
    \includegraphics[width=0.33\linewidth]{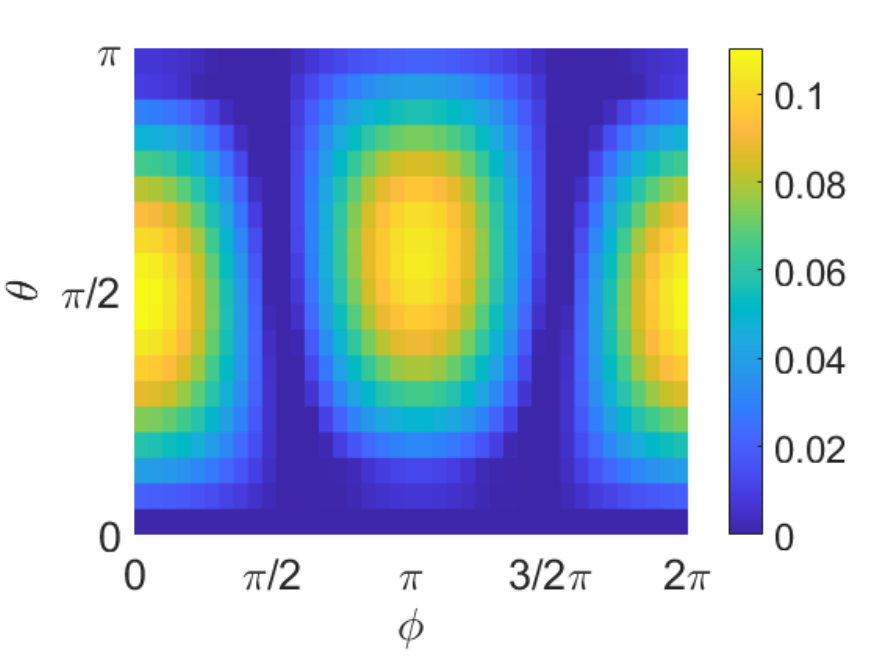}
    \includegraphics[width=0.33\linewidth]{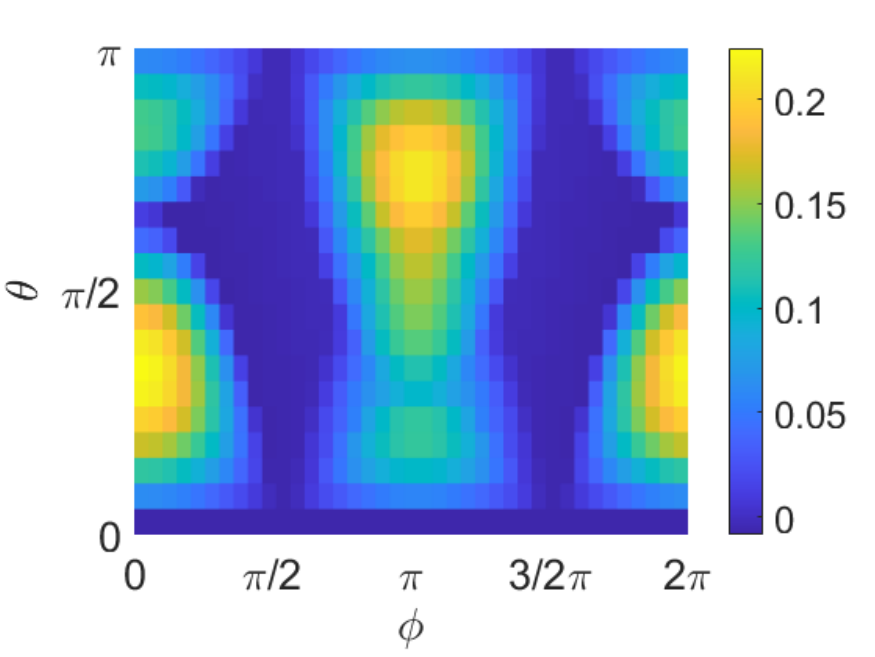}
    \caption{Negativity of the MHQ distribution of the work done by the defect, for the initial state of \eqref{eq:Bloch_state}. From left to right, the value of the perturbation strength increases as $k=0.1,1,100$.}
    \label{fig:Bloch_analysis_negativity}
\end{figure*}

In Fig.~\ref{fig:LoschmidtEcho_BlochState_various_k} we analyze $|\nu(t)|^2$ considering the two-level initial state of Eq.~\ref{eq:Bloch_state}, for increasing perturbation strengths $k = 0.1$ [panel (a)], $k = 1$ [panel (b)], and $k = 100$ [panel (c)]. In each panel, we vary the parameters $\theta$ and $\phi$ with the aim to explore how different initial superpositions affect the extent and rate of orthogonalization. The value of $|\nu(t)|^2$ for $\theta=0$ (blue dashed line), corresponding to the ground state, is essentially constant in the weak perturbation regime ($k=0.1$ and $k=1$), whereas it exhibits a periodic revival with period $t=\pi$ for strong perturbation ($k=100$). In the weak perturbation regime [panels (a)-(b)], an equal-weight superposition ($\theta=\pi/2$) leads to a complete decay of $|\nu(t)|^2$, independently of the phase $\phi$. On the contrary, with strong perturbations [panel (c)], the maximum orthogonalization occurs for $\theta=\pi$, which corresponds to initialize the particle in the second excited state. For lower values of $k$, setting $\theta = 0$ and $\theta = \pi$ produces nearly identical dynamics. The phase $\phi$ generally has a marginal influence; however, for $k=100$ and $\theta=\pi/2$, varying $\phi$ causes $|\nu(t)|^2$ to be mirrored with respect to the midpoint of the revival period. As $k$ increases, the revival period of the LE decreases and eventually tends to the value $\pi$. These behaviours reflect the growing impact of the perturbation on the system’s dynamics and highlight the crucial role of quantum coherence in modulating the response of the LE. 

\begin{figure}[b]
    \centering
    \includegraphics[width=0.85\linewidth]{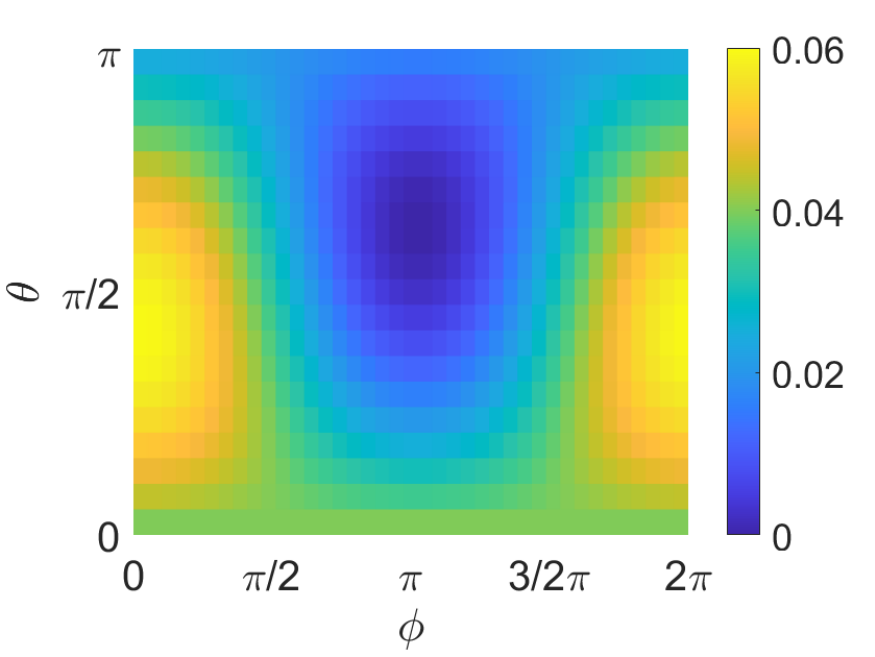}
    \caption{Average $\langle w\rangle$ of the KDQ distribution of work, calculated using the initial state of \eqref{eq:Bloch_state} for different angles $\theta$ and $\phi$ with $k=0.1$. 
    }
    \label{fig:Bloch_analysis_MeanWork}
\end{figure}

We now analyze the non-classical features introduced by the delta-perturbation at the level of the negativity of the real part of the KDQ distribution, also known as Margenau-Hill quasiprobability (MHQ) distribution~\cite{margenau1961correlation,allahverdyan2014nonequilibrium,lostaglio2018quantum,diaz2020quantum,hernandez2022experimental,PeiPRE2023}. To do this, we use the functional~\cite{GherardiniTutorial} 
\begin{equation}\label{eq:nonpositivity}
    \mathcal{N}_{\rm Re} \equiv -1 + \sum_{m,n} \left| {\rm Re}\,q_{m,n} \right|,
\end{equation}
which quantifies the departure of the distribution from the classical probabilistic behaviour following the Kolmogorov axioms. Fig.~\ref{fig:Bloch_analysis_negativity} shows $\mathcal{N}_{\rm Re}$ as a function of $\theta$ and $\phi$ for increasing perturbation strengths: $k = 0.1$ (left), $k = 1$ (center), and $k = 100$ (right). At the poles of the Bloch sphere ($\theta = 0$ and $\theta = \pi$), corresponding to the eigenstates of $\hat{H}_i$, the distribution of work is positive and $\mathcal{N}_{\rm Re}$ vanishes. Instead, in the weak perturbation regime ($k = 0.1$), the negativity of the distribution is maximal along the equator of the Bloch sphere ($\theta = \pi/2$), where the initial state is in an equally-weighted superposition of the eigenstates. As $k$ increases, $\mathcal{N}_{\rm Re}$ develops pronounced interference-like patterns across the $\theta$-$\phi$ plane, with two main lobes of high negativity. This indicates that not only the superposition amplitude (controlled by $\theta$) but also the relative phase $\phi$, entering the initial state, plays a significant role in determining the dynamical response of the system to the delta-perturbation.

The average work $\langle w\rangle$ done by the defect on the system after the quench transformation contains only two contributions given the choice of the initial state; it can be computed from the KDQ distribution as
\begin{equation}
    \langle w\rangle = \sum_{m=0}^\infty q_{0,m} \Delta E_{m,0} + \sum_{m=0}^\infty q_{1,m} \Delta E_{m,1}, 
\end{equation}
where $\Delta E_{m,n} = E_m' - E_n$ is the energy difference between the eigenstates of the perturbed and unperturbed Hamiltonians. 

\begin{figure}[b]
    \centering
    \includegraphics[width=0.85\linewidth]{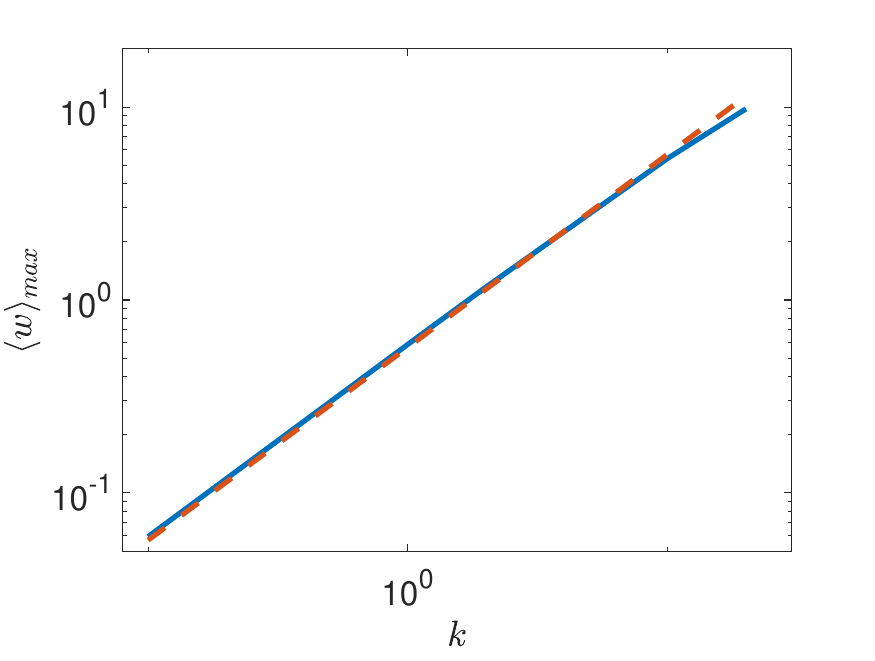}
    \caption{Linear increase of the maximum average work $\langle w \rangle_{max}$ as a function of $k$ (blue solid line). The red dashed line corresponds to the fit $\langle w \rangle_{max}\cong0.57 \, k$ }
    \label{fig:Bloch_analysis_MeanWork_max}
\end{figure}

In Fig.~\ref{fig:Bloch_analysis_MeanWork}, we plot the dependence of $\langle w\rangle$ on the angles $\theta$ and $\phi$, for $k=0.1$. The average work exhibits a characteristic structure with one maximum located around the equator ($\theta = \pi/2$) at $\phi=0$. This maximum corresponds to initial states that are equally-weighted coherent superpositions of the first two eigenstates of $\hat{H}_i$. This structure remains unchanged for different values of the perturbation strength $k$, though the values assumed by $\langle w \rangle$ change with $k$. In this regard, defining $\langle w\rangle_{\rm max}$ as the maximum value of $\langle w \rangle$, in Fig.~\ref{fig:Bloch_analysis_MeanWork_max} we show that $\langle w\rangle_{\rm max}$ increases linearly with the perturbation strength $k$.

\begin{figure*}[th]
    \centering
    \includegraphics[width=0.53\linewidth]{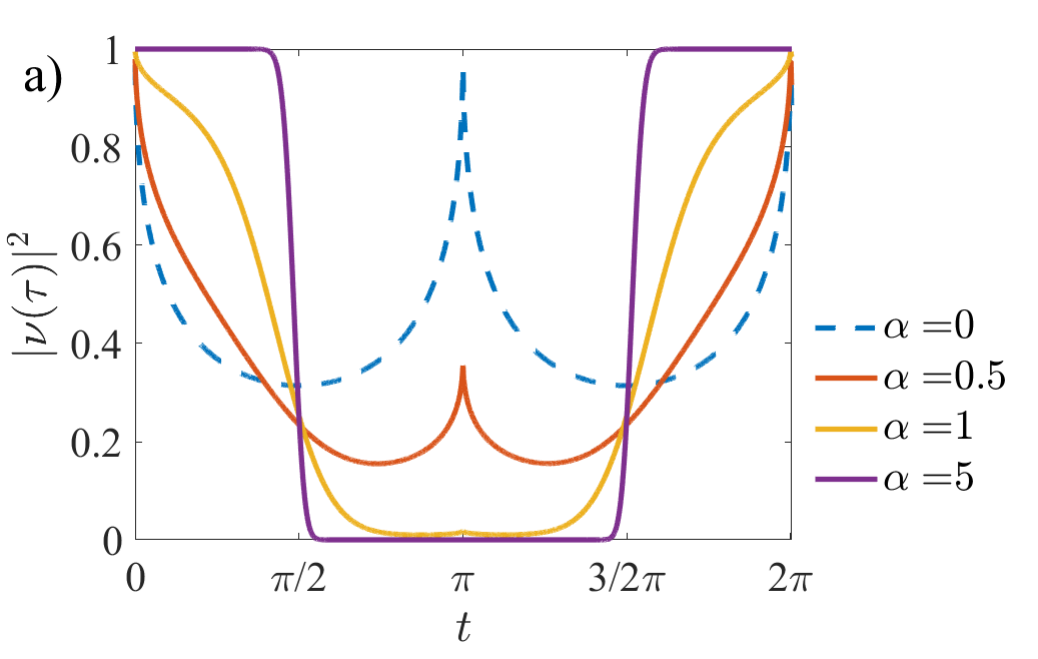}
    \includegraphics[width=0.45\linewidth]{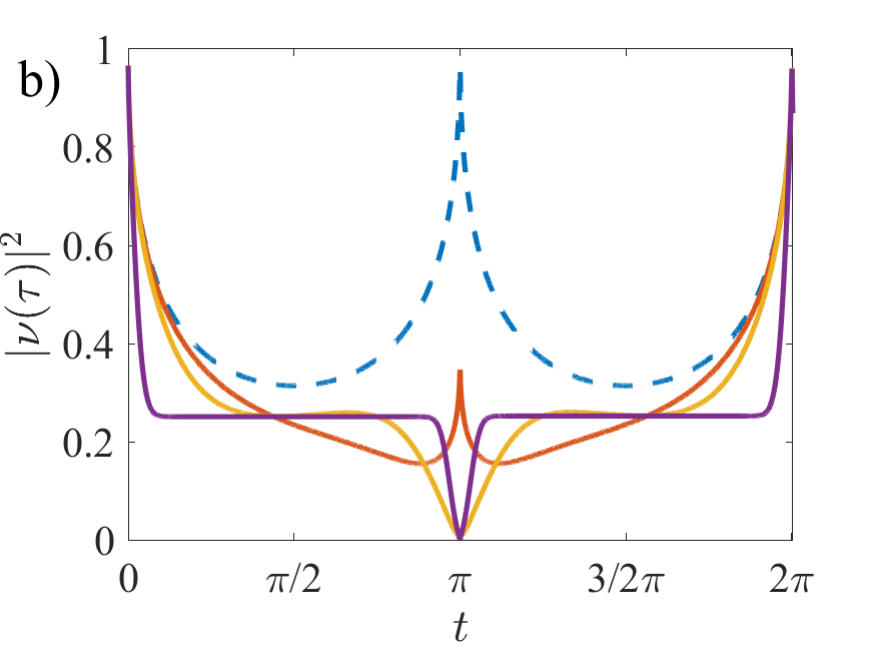} \\
    \includegraphics[width=0.33\linewidth]{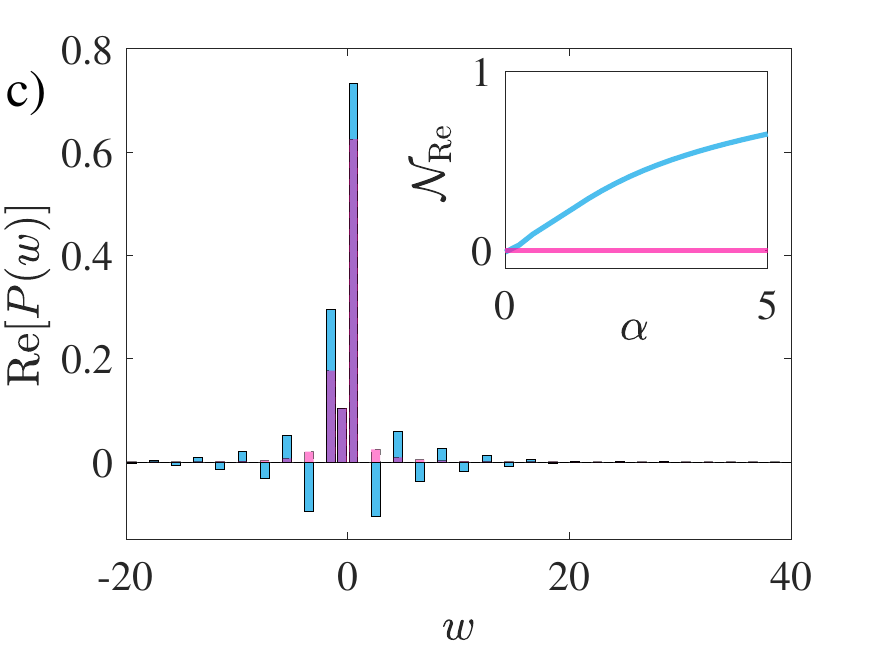}
    \includegraphics[width=0.33\linewidth]{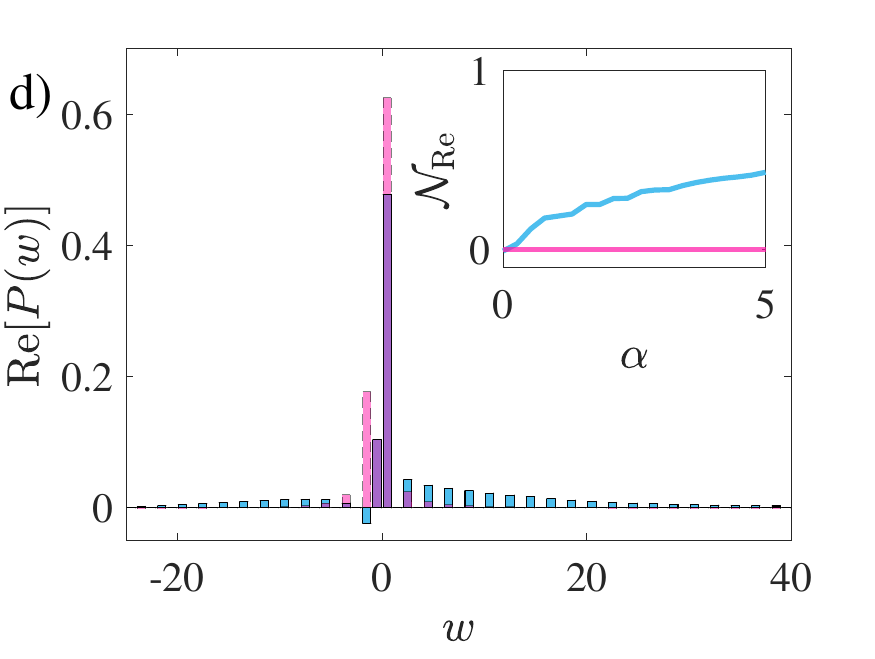}
    \includegraphics[width=0.33\linewidth]{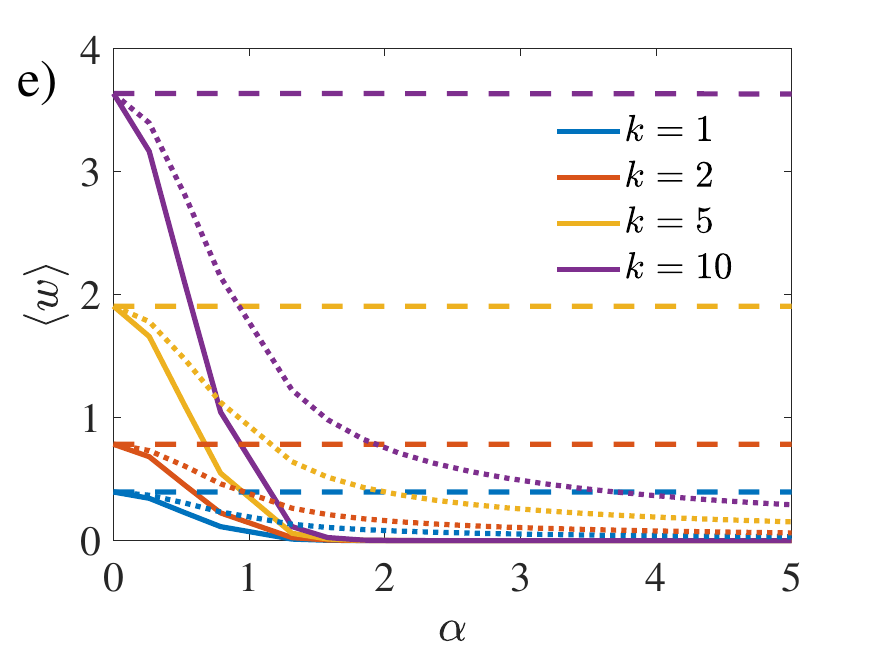}
    \caption{(a-b) Time evolution of $|\nu(t)|^2$ for a particle initialized in the coherent state of \eqref{eq:CoherentState}, with real amplitude $\alpha$ ranging from $0$ (ground state, represented by the dashed blue curve) to $\alpha = 5$, with perturbation strength $k = 10^3$. Notice that, differently to panel (a), in panel (b) the initial state includes also an alternating phase factor $(-1)^{n/2}$. (c-d-e) Analysis of the work distribution and corresponding average work. (c) MHQ distribution of work for an initial coherent state with $\alpha = 5$ (light blue). We compare it with the distribution obtained by initializing the particle in the diagonal state (pink) associated to the coherent state whose off-diagonal elements are set to zero. Inset: Non-positivity of the distribution (in terms of negativity quantified by $\mathcal{N}_{\rm Re}$) as a function of $\alpha$, for the coherent state (light blue) and the corresponding diagonal state (pink). 
    (d) Same plot as in panel (c), with the only difference that the initial state includes also the alternating phase factor $(-1)^{n/2}$. When the particle is initialized in the coherent state, the work distribution shows a marked discontinuity around $w=0$ and lower values of $\mathcal{N}_{\rm Re}$.
    (e) Average work $\langle w\rangle$ as a function of $\alpha$ for the interaction strengths $k=1,2,5,10$. 
    Initializing the particle in the coherent (solid lines) and diagonal (dotted lines) states leads to a decreasing value of the mean work by increasing $\alpha$, while the initial coherent state including the phase factor (dashed lines) entails that $\langle w\rangle$ is practically flat. 
    }
    \label{fig:Coherent_state_analysis}
\end{figure*}

\subsection{Coherent state}

Let us initialize the particle in a coherent state, which offers more structured quantum coherence patterns beyond the case of a superposition of only two states. Coherent states are defined as
\begin{equation}\label{eq:CoherentState}
    \ket{\alpha} = e^{-|\alpha|^2/2} \sum_{n=0}^{\infty} \frac{\alpha^n}{\sqrt{n!}} \ket{n},
\end{equation}
where $\ket{n}$ are the eigenstates of the unperturbed Hamiltonian $\hat{H}_i$ and $\alpha \in \mathbb{C}$ is the coherent amplitude. In numerical simulations, the sum is truncated at a cut-off $N_{\rm max}$ sufficiently large to ensure convergence. These states are widely recognized for allowing control of both the amplitude and the phase of the initial state.

Fig.~\ref{fig:Coherent_state_analysis}(a)-(b) illustrate the time evolution of $|\nu(t)|^2$ in the strong perturbation regime ($k = 10^3$), for initial coherent states with different values of $\alpha$ and phases. In Fig.~\ref{fig:Coherent_state_analysis}(a), for $\alpha=0$ (representing the ground state, plotted with the dashed blue line), the dynamics exhibit perfect periodicity with revival period $T = \pi$, and $|\nu(t)|^2$ always takes a value quite larger than zero, which indicates a lack of orthogonalization. However, as $\alpha$ increases (solid lines), we observe a progressive flattening of $|\nu(t)|^2$ around $t = \pi$, while maintaining the revival at a value close to $1$ for $T=2\pi$.

Fig.~\ref{fig:Coherent_state_analysis}(b) presents the time evolution of $|\nu(t)|^2$ for initial coherent states that include also an additional alternating phase factor $(-1)^{n/2}$ applied to the coefficients. This ``fermion-like'' phase factor can be considered when mimicking an antisymmetric many-body fermionic wave-function even in the single-particle case~\cite{paper_short}. At the level of the overlaps calculation (refer to Fig.~\ref{fig:Lambda_m0_m2_m20}), the phase factor $(-1)^{n/2}$ allows to obtain ordered zones of positivity and negativity in the work distribution. This reflects in different behaviours of $|\nu(t)|^2$, which is still maintained flattened but at a non-zero value, and exhibits a sharp dip to zero precisely at $t = \pi$.

To understand the influence of the perturbation strength $k$, we investigate $|\nu(t)|^2$ for different values of $k$. As $k$ is decreased, the flattening of $|\nu(t)|^2$ observed at $t = \pi$ is progressively suppressed, and the revival times shift to later values. This trend mirrors the behaviour previously found in the analysis of the case of a superposition of two states (see Fig.~\ref{fig:LoschmidtEcho_BlochState_various_k}), confirming the generality of the $\pi$-periodicity in the strong quench regime ($k$ large) and the sensitivity of the orthogonalization to both the perturbation strength and the coherent-phase structure of the initial state.

As discussed in Sec.~\ref{sec:LE_Pw}, the Fourier transform of the LE yields the KDQ distribution of work. In this regard, in Fig.~\ref{fig:Coherent_state_analysis}(c) we present the MHQ work distributions for a particle initially prepared in a coherent state with amplitude $\alpha = 5$, represented by the light blue histogram. It is compared with the pink histogram associated to the scenario where the initial state is diagonal, given by removing all the off-diagonal elements in the $\hat{H}_i$ basis. The primary distinction in the histograms lies in their widths: the inclusion of quantum coherences in the initial state leads to a significantly greater energy spread, which indicates a delocalization in energy space. The inset in Fig.~\ref{fig:Coherent_state_analysis}(c) illustrates the functional $\mathcal{N}_{\rm Re}$ of \eqref{eq:nonpositivity}, which increases with the coherence amplitude $\alpha$, as in the simpler case with the states defined in \eqref{eq:ket_Psi0}. 

\begin{figure*}[bt]
    \centering
    \includegraphics[width=0.53\linewidth]{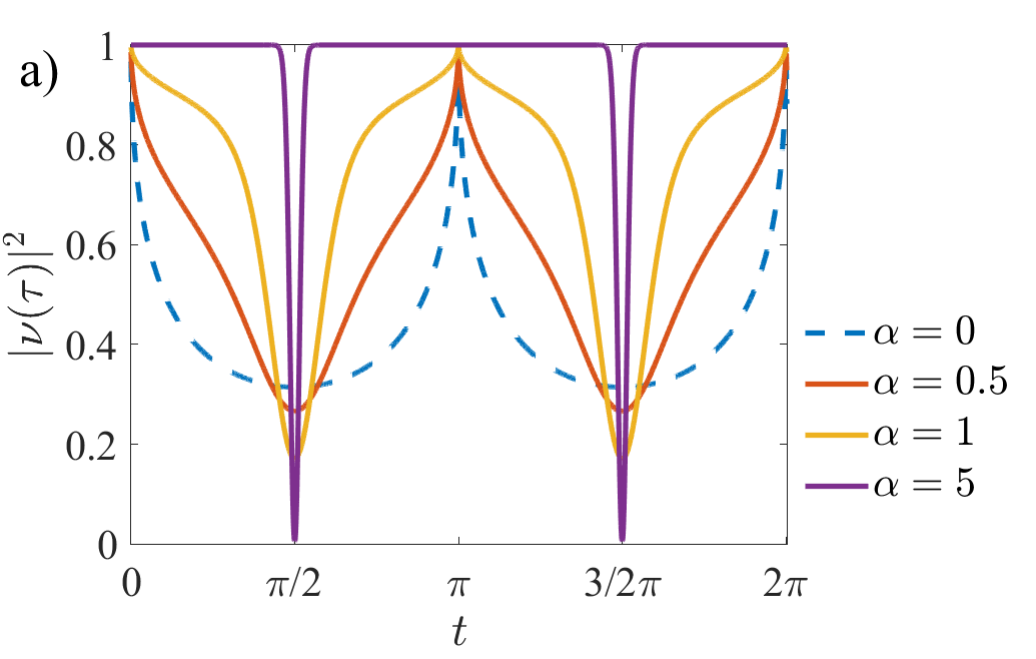}
    \includegraphics[width=0.45\linewidth]{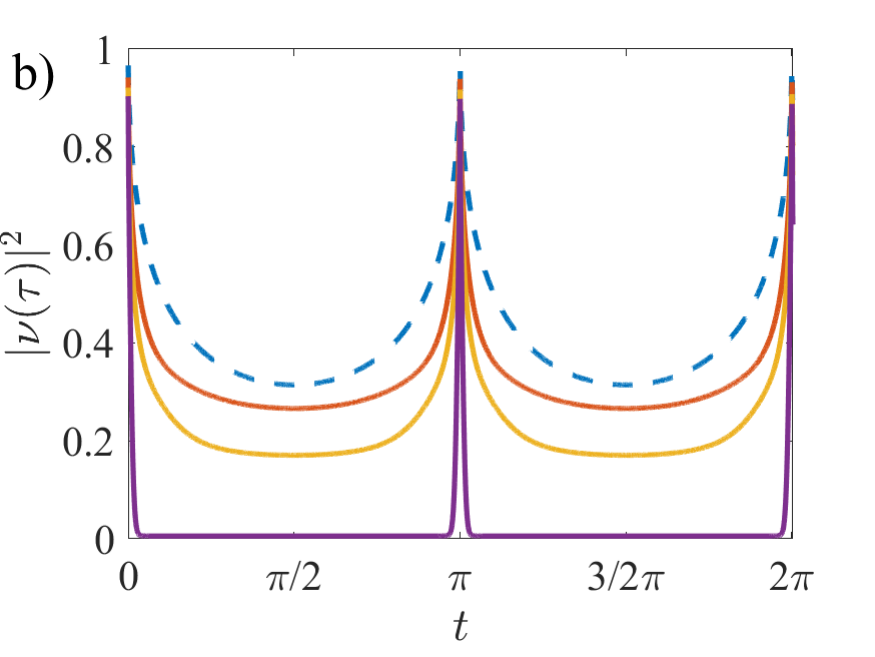} \\
    \includegraphics[width=0.33\linewidth]{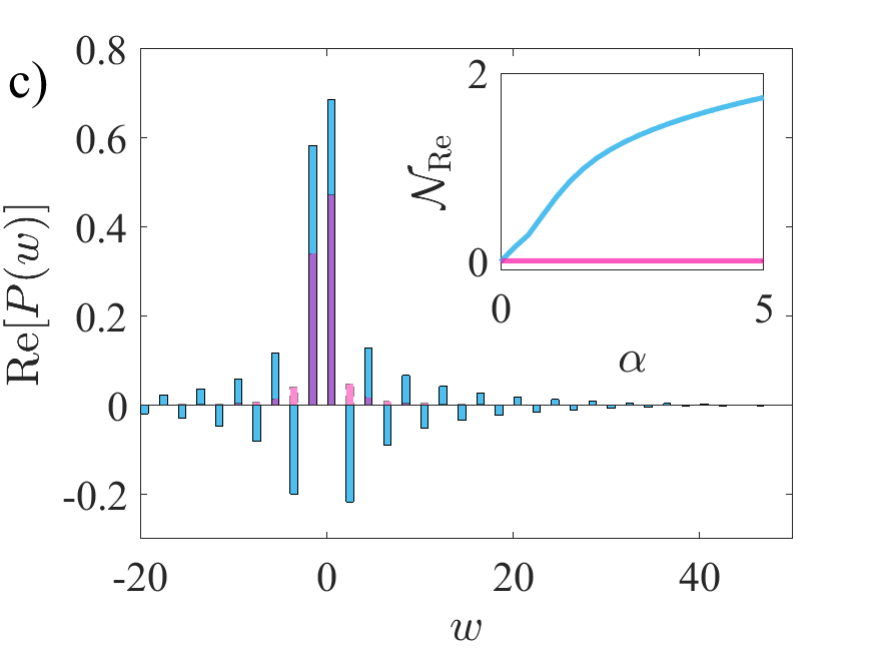}
    \includegraphics[width=0.33\linewidth]{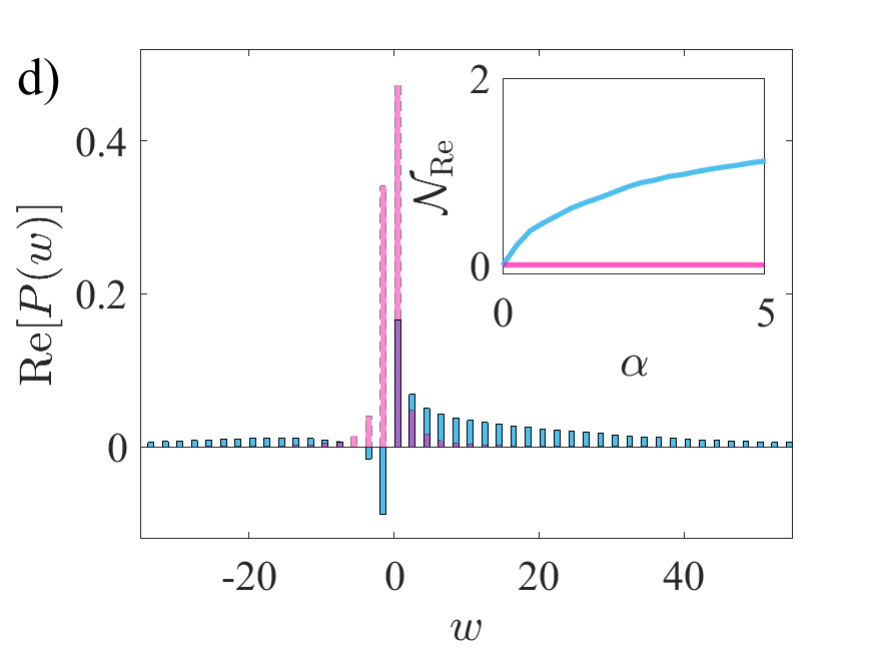}
    \includegraphics[width=0.33\linewidth]{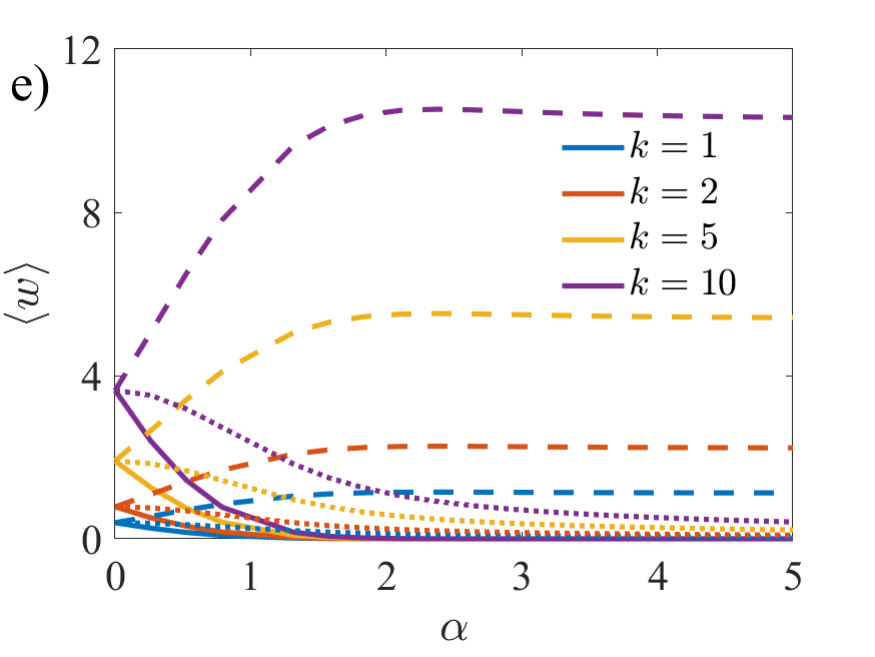}
    \caption{(a-b) Time evolution of $|\nu(t)|^2$ for a particle initialized in the cat state of \eqref{eq:kitten_state}, perturbed by a localized defect with strength $k = 10^3$. 
    (a) Standard cat state with coherence amplitude $\alpha$ from $0$ (ground state, dashed blue) to $5$: $|\nu(t)|^2$ is close to unity, with sharp dips at half-periods, which indicates no orthogonalization for most of the time. 
    (b) Analogous initial states as in (a) but with the inclusion of the alternating phase factor $(-1)^{n/2}$. 
    (c) Work MHQ distribution for the cat state with $\alpha = 5$ (light blue) and its diagonal counterpart (pink). The broader spread of the distribution considering an initial cat state means a stronger energy-space delocalization. Inset: Non-positivity functional $\mathcal{N}_{\rm Re}$ as a function of $\alpha$.
    (d) Work MHQ distribution due to initializing the particle in a cat state whose coefficients also include the phase factor $(-1)^{n/2}$. 
    (e) Average work $\langle w \rangle$ as a function of the coherence amplitude $\alpha$ for various interaction strengths $k$. Initial cat states (solid lines) and corresponding diagonal states (dotted lines) leads to a $\langle w \rangle$ decreasing with $\alpha$. Dashed lines, instead, denote the case of initial cat states with the alternating phase factor, whereby $\langle w \rangle$ increases and saturates with $\alpha$. 
    }
    \label{fig:Kitten_state_LE_k1000}
\end{figure*}

Fig.~\ref{fig:Coherent_state_analysis}(d) shows the MHQ distributions of work for analogous initial states of panel (c) but including also the ``fermion-like'' phase factor $(-1)^{n/2}$. The presence of coherences in the initial state (light blue) again leads to a broader distribution of work compared to the incoherent case (pink). 
Interestingly, in the former case, a discontinuity emerges in the distribution, which is a signature of quantum superposition effects, as discussed earlier in Sec.~\ref{eq:overlaps_strong_k}. Furthermore, the non-positivity functional $\mathcal{N}_{\rm Re}$ reaches higher values in the absence of this phase factor, since many negative instances of $P[w]$ (see \eqref{eq:work_distribution}) in panel (c) are positive in such a case.

Finally, Fig.~\ref{fig:Coherent_state_analysis}(e) displays the average work $\langle w \rangle$ done on the system by the delta-perturbation, as a function of the coherence amplitude $\alpha$ for various interaction strengths $k$. For the cases with initial coherent (solid lines) and diagonal (dotted lines) states, $\langle w \rangle$ increases with $\alpha$ and eventually saturates. Conversely, including the phase factor $(-1)^{n/2}$ in the coefficients of the initial coherent state makes the average work practically constant for any $\alpha$ (dashed lines).

\subsection{Cat state}

We now consider a class of highly non-classical states that extend beyond coherent states, specifically referred to as cat states. These states are symmetric superpositions of coherent states with opposite phases and exemplify quantum interferences at the mesoscopic scale. In particular, we concentrate on the even Schr\"{o}dinger cat state, which is proportional to
\begin{equation}\label{eq:kitten_state}
    \ket{\SchrodingersCat{0}} \propto \ket{\alpha}+\ket{-\alpha}.
\end{equation}
The name of $\ket{\SchrodingersCat{0}}$ is justified by the fact that it can be viewed as the projection of a coherent state onto the even subspace of the system's Hilbert space. The non-classical nature of this state becomes evident in its Wigner function (see Fig.~\ref{fig:Wigner}e), which --- unlike the Wigner function of a coherent state --- exhibits an interference pattern characterized by regions of negativity. These negative values indicates non-classical correlations naturally present in $\ket{\SchrodingersCat{0}}$. In addition to these correlations, the quench with a delta-perturbation further induces non-classical features, as revealed by the time evolution of the Wigner function of the system's state, shown in Fig.~\ref{fig:Wigner}(a-b).

In Fig.~\ref{fig:Kitten_state_LE_k1000}(a), we analyze the time evolution of $|\nu(t)|^2$ for increasing values of $\alpha$ in the strong quench regime with $k=10^3$. 
As before, the dashed blue line corresponds to the case of initializing the particle in the ground state ($\alpha=0$), which is plotted for reference. We observe that for $\alpha=5$ $|\nu(t)|^2$ remains close to unity for any time $t$, showing sudden periodical dips at $t = (j+1)\pi/2$ with $j$ integer. This behaviour indicates that no orthogonalization occurs despite the presence of strong perturbations.

For an initial cat state, the effects of a delta-perturbation on the response of $|\nu(t)|^2$ are significantly pronounced if the ``fermion-like'' phase factor $(-1)^{n/2}$ is included in the initial superposition~\cite{paper_short}. This is shown in Fig.~\ref{fig:Kitten_state_LE_k1000}(b) where we observe that the system orthogonalizes most rapidly. Indeed, $|\nu(t)|^2$ decays to zero almost immediately and remains to zero for a longer time interval, while maintaining the periodic revival at $1$ with period $T=\pi$. Increasing $\alpha$, this trend is enhanced, thus emphasising the crucial role of phases in controlling the dynamics.

The MHQ distribution of work, due to the quench transformation, is plotted in panels (c)-(d) of Fig.~\ref{fig:Kitten_state_LE_k1000}. The initialization in a cat state (light blue) entails a broader and more structured distribution compared to its diagonal counterparts (pink) associated with the corresponding initial diagonal state. The inclusion of the phase factor $(-1)^{n/2}$ again results in a visibly sharper structure of the distribution that reveals discontinuities.

Finally, the panel Fig.~\ref{fig:Kitten_state_LE_k1000}(e) shows the average work $\langle w\rangle$ as a function of $\alpha$, for different values of $k$. The results for both the superposition case (solid lines) and the diagonal counterpart (dotted lines) decreases with $\alpha$. The most relevant finding is the one with the phase factor, where $\langle w\rangle$ increases with $\alpha$ until saturation (dashed lines). 

\begin{figure*}[t]
    \centering
    \includegraphics[width=0.42\linewidth]{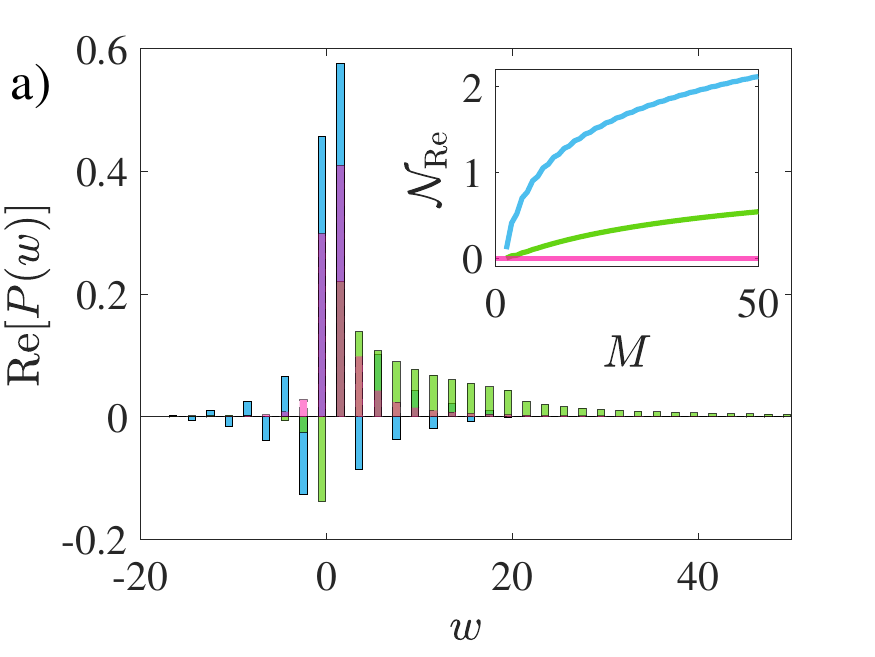}
    \includegraphics[width=0.42\linewidth]{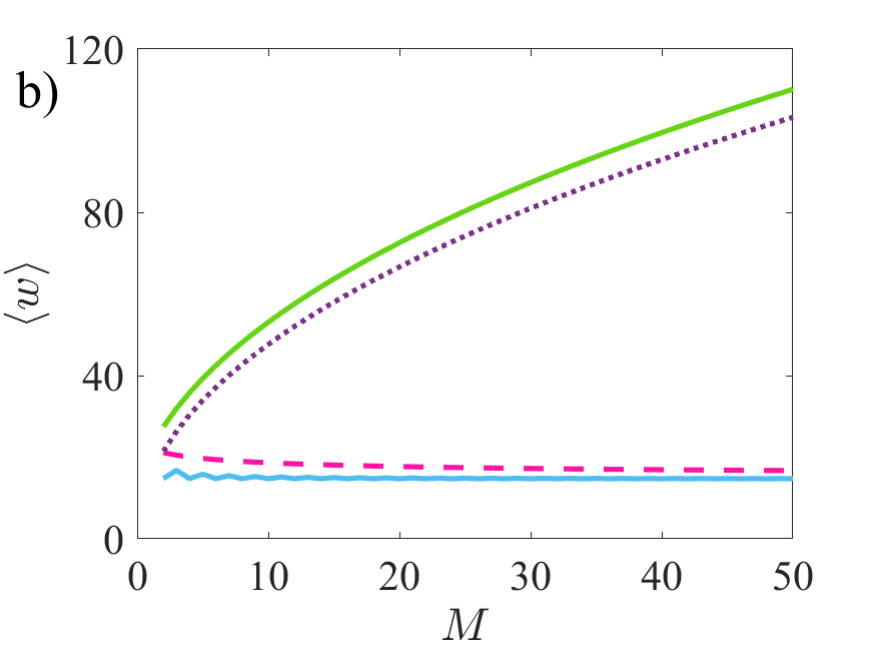}
    \caption{(a) Work MHQ distribution for two-fermions initialized in the superposition state of \eqref{eq:TwoFermionState}, without (light blue) and with (green) the alternating phase factor $(-1)^{n/2}$ multiplying the coefficients of the superposition. The corresponding diagonal state is depicted in pink. 
    (b) Average work $\langle w \rangle$ as a function of the superposition size $M$.  Results are shown for the two-fermions whose initial state is without (light blue) and with (green) the alternating phase factor, as well as for the initial diagonal state (dashed pink line). For comparison, we also show the average work for the single-particle case with the initial state containing the phase factor $(-1)^{n/2}$ (dotted purple line). 
    }
    \label{fig:TwoFermions_q_hist}
\end{figure*}

\section{Case-study 2: Two fermions}\label{sec:two_fermions}

Let us extend the analysis to the scenario with two non interacting fermions ($N=2$), subjected to a delta-perturbation. We consider an initial pure state that in general does not commute with the initial Hamiltonian of the fermionic system, i.e.,
\begin{equation}\label{eq:initial_state_2-fermions}
    \ket{\Psi(0)} = \sum_{k_1 < k_2} \alpha_{k_1,k_2} \ket{k_1,k_2}_{\rm Slater},
\end{equation}
where $\ket{k_1,k_2}$ is the properly anti-symmetrized Slater determinant built from the single-particle eigenstates $\ket{k_1}$ and $\ket{k_2}$ of the unperturbed Hamiltonian $\hat{H}_i$. In other words, $\ket{\Psi_{\rm Slater}(0)}$ of \eqref{eq:initial_state_2-fermions} is a many-body wave-function whose projection on the position space follows \eqref{eq:proj_many-body_Slater}. 
The sum is restricted to $k_1 < k_2$ in order to avoid redundancy due to antisymmetry.

Following the same procedure used for \eqref{eq:LE} in the single-particle case, we can rewrite the LE as
\begin{widetext}
\begin{eqnarray}\label{LE_2fermions}
    \nu(t) &=& {\rm Tr}\left[ \hat{U}'(t) \ketbra{\Psi(0)}{\Psi(0)} \hat{U}(-t) \right] = \nonumber \\ 
    &=& \sum_{n_1<n_2} \bra{n_1,n_2} e^{-i\hat{H}'_St} \left( \sum_{m_1'<m_2'} \ketbra{m_1',m_2'}{m_1',m_2'} \right) \left( \sum_{k_1<k_2} \sum_{l_1<l_2} \alpha_{k_1,k_2} \alpha_{l_1,l_2}^* \ketbra{k_1,k_2}{l_1,l_2} \right) e^{i\hat{H}_S t} \ket{n_1,n_2} = \nonumber \\
    &=& \sum_{m_1'<m_2'} \sum_{n_1<n_2} e^{-i(E_{m_1'}'+E_{m_2'}'-E_{n_1}-E_{n_2})t} \Lambda_{n_1,n_2,m_1',m_2'} \alpha_{n_1,n_2}^*  \sum_{k_1<k_2} \alpha_{k_1,k_2} \Lambda_{m_1',m_2',k_1,k_2},
\end{eqnarray}    
\end{widetext}
where the superscript $(')$ refers to the perturbed eigenstates, and we have defined the many-body overlaps between the anti-symmetrized perturbed and unperturbed two-particle eigenstates as $\Lambda_{m_1' m_2',n_1 n_2} \equiv \langle m_1',m_2'|n_1,n_2\rangle$. According to the approach outlined in Sec.~\ref{sec:overlap_many-body_state}, the overlap between the many-body wave-functions can be computed using the determinant of the matrix whose elements are given by the single-particle overlaps. For a system made of two fermions, these overlaps reduce to compute the determinant
\begin{equation}
    \Lambda_{m_1' m_2',n_1 n_2} = 
    \begin{vmatrix}
        \Lambda_{m_1',n_1} & \Lambda_{m_1',n_2} \\
        \Lambda_{m_2',n_1} & \Lambda_{m_2',n_2}
    \end{vmatrix} \,,
\end{equation}
where $\Lambda_{m,n} = \langle m'|n\rangle$ is the single-particle overlap. Recalling that the LE is the characteristic function of a KDQ distribution, \eqref{eq:LE}, one has that
\begin{equation}
    \nu(t) = \sum_{m_1',m_2'} \sum_{n_1,n_2} e^{-i(E_{m_1'}'+E_{m_2'}'-E_{n_1}-E_{n_2})t} q_{n_1,n_2,m_1',m_2'} \,,
\end{equation}
where $q_{n_1,n_2,m_1',m_2'}$ is the two-particle KDQ. 
This quasiprobability can be explicitly expressed as 
\begin{equation}\label{eq:KDQ_2fermions}
    q_{n_1,n_2,m_1',m_2'} \equiv \alpha_{n_1 n_2}^* \Lambda_{m_1' m_2',n_1 n_2}^* \sum_{k_1,k_2} \alpha_{k_1,k_2} \Lambda_{m_1' m_2',k_1 k_2} 
\end{equation}
that directly extends the single-particle result previously discussed in \eqref{eq:q_mn}.

Let us assume the two fermions to be prepared in a uniform superposition of $M$ anti-symmetrized eigenstates. In such configuration, all coefficients are equal to $\alpha = 1/\sqrt{M}$ and the corresponding KDQ takes the form
\begin{eqnarray}
    &&
    q_{n_1,n_2,m_1',m_2'} =\nonumber\\ 
    &&
    = \frac{1}{M}
    \begin{vmatrix}
        \Lambda_{m_1',n_1} & \Lambda_{m_1',n_2} \\
        \Lambda_{m_2',n_1} & \Lambda_{m_2',n_2}
    \end{vmatrix}
    \sum_{k_1<k_2}
    \begin{vmatrix}
        \Lambda_{m_1',k_1} & \Lambda_{m_1',k_2} \\
        \Lambda_{m_2',k_1} & \Lambda_{m_2',k_2}
    \end{vmatrix}.
\end{eqnarray}
More specifically, we consider an initial state of the form 
\begin{equation}\label{eq:TwoFermionState}
    \ket{\Psi(0)} = \frac{1}{\sqrt M}\sum_{\substack{n=2 \\ n \, \mathrm{even}}}^M \ket{0,\,n}\,,
\end{equation}
where the sum starts from $n=2$ to respect the Pauli exclusion principle by avoiding double occupancy of the single-particle ground state. Notice that the corresponding anti-symmetrized wave-function of \eqref{eq:TwoFermionState} is $\ket{\Psi_{\rm Slater}(0)} = \sum_{n=2}^M \frac{1}{\sqrt{2M}} \left(\ket{0,\,n}-\ket{n,\,0}\right)$. Choosing \eqref{eq:TwoFermionState} enables us to systematically explore how the size of the superposition $M$ influence the response of the system (here, two fermions) that exhibits coherence-enhanced orthogonalization under a delta-perturbation.

In order to investigate the role of quantum phases in the initial state, we also examine a modified version of $\ket{\Psi(0)}$ that also includes the alternating phase factor $(-1)^{n/2}$: 
\begin{equation}\label{eq:TwoFermionState_menouno}
    \ket{\Psi(0)} = \sum_{\substack{n=2 \\ n \, \mathrm{even}}}^M (-1)^{n/2} \frac{\ket{0\,n}}{\sqrt{M}}\,.
\end{equation}

Fig.~\ref{fig:TwoFermions_q_hist}(a) shows the work MHQ distributions for both the initial superposition state in \eqref{eq:TwoFermionState} (light blue) and its counterpart of \eqref{eq:TwoFermionState_menouno} with the alternating phase factor (green). Interestingly, the histogram resembles the single-particle case, but with the addition of a jump in the long tail of the distribution for positive $w$. This trend might be not universal but dependent on the specific choice of the initial state as the one of \eqref{eq:TwoFermionState_menouno}.  
For comparison, we plot also the histogram of their corresponding diagonal state (pink), which is always positive. This fact is quantified in the inset, where the non-positivity of the distribution, measured by $\mathcal{N}_{\rm Re}$, is plotted. The emergence of negativity due to coherence mirrors the single-particle case \cite{paper_short}. 
The same colour scheme is used in Fig.~\ref{fig:TwoFermions_q_hist}(b) that displays the average work $\langle w\rangle$ as a function of the size of the superposition $M$. The average work is nearly constant for both the initial coherent superposition without the phase $(-1)^{n/2}$ (light blue) and the diagonal state (dashed pink line). In contrast, when the phase factor is included in the initial state (green), $\langle w\rangle$ increases with $M$, thus indicating an enhancement of the energy transfer induced by the quench transformation. For comparison, in Fig.~\ref{fig:TwoFermions_q_hist}(b) we plot the corresponding average work for a single particle whose state includes $(-1)^{n/2}$ (dotted purple line). We observe that this initial state yields nearly the same average work of the two-fermion case with the same phase factor. This is due to the fact that the initial energy of one of two fermions is kept at zero, being in the ground state. In the more general case, where the initial energies of the two fermions take arbitrary values, we expect a linear dependence of $\langle w\rangle$ with the number of fermions. 

\begin{figure*}
    \centering
    \begin{minipage}{0.38\linewidth}
    \includegraphics[width=\linewidth]{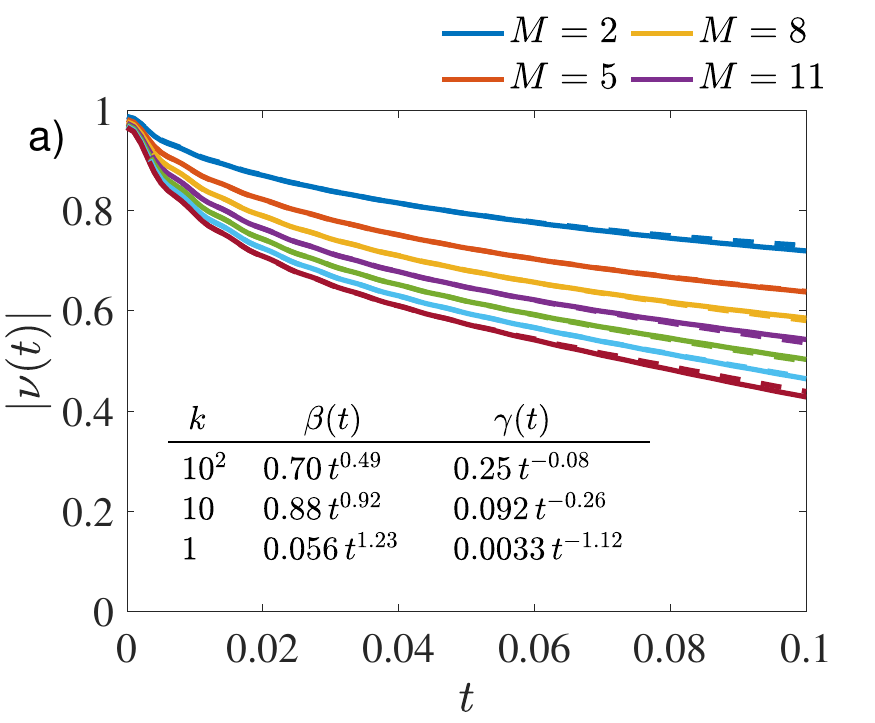}
    \end{minipage}
    \hspace{-10pt}
    \begin{minipage}{0.38\linewidth}
    \includegraphics[width=\linewidth]{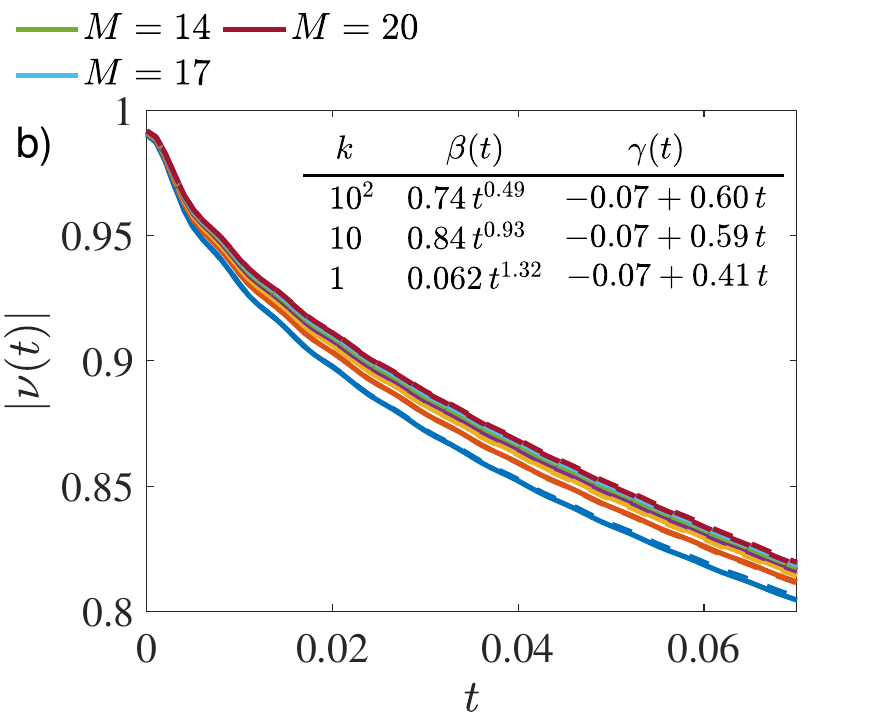}
    \end{minipage}
    \hspace{-10pt}
    \begin{minipage}{0.23\linewidth}
    \includegraphics[width=\linewidth]{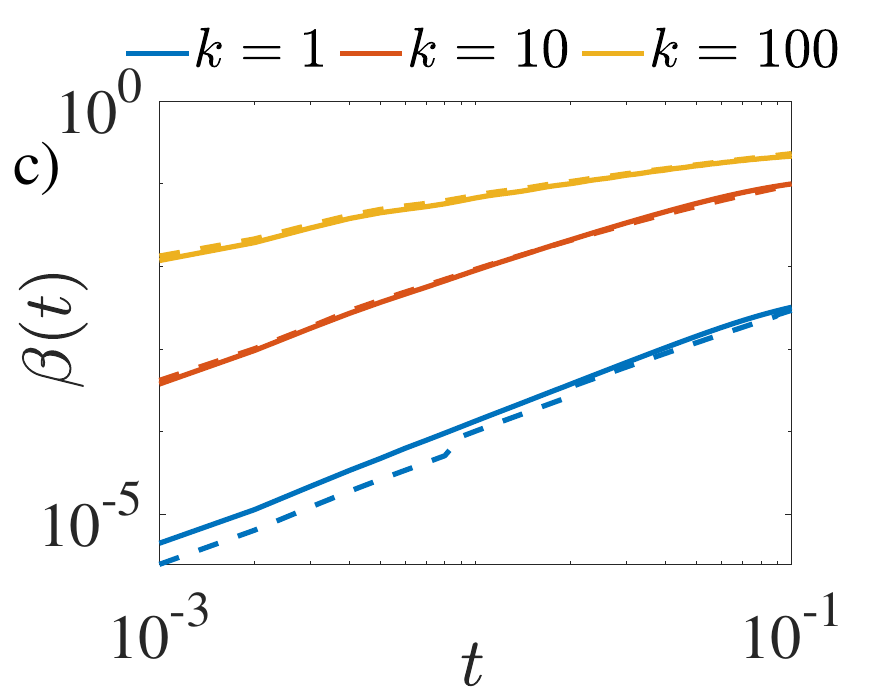}
    \includegraphics[width=\linewidth]{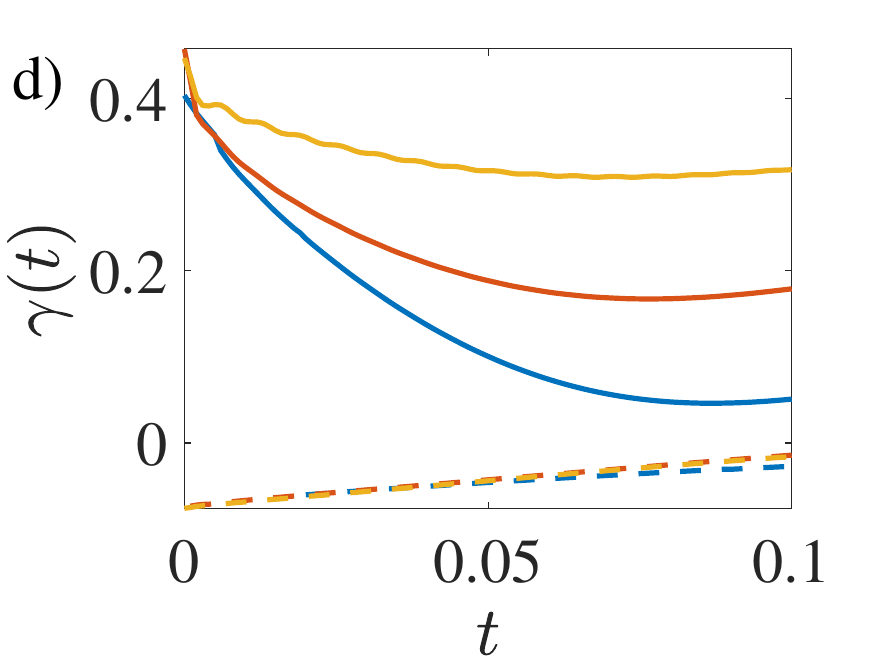}
    \end{minipage}
    \caption{Time evolution the LE modulus for two fermions subjected to a delta perturbation. 
    (a)-(b) $|\nu(t)|$ as a function of time, for two fermions initialized in a coherent anti-symmetrized superposition of energy eigenstates [panel (a)] and in the corresponding incoherent diagonal state [panel (b)], respectively. Here, the fermions are perturbed by a delta-function potential of strength $k=100$. Solid lines represent the numerical results, while dashed lines refer to the fitted curves following \eqref{eq:LE_fit}. 
    Insets in panels (a)-(b): fitted decay coefficients $\beta(t)$ and $\gamma(t)$ as a function of time. 
    Panels (c)-(d): time-dependence of $\beta(t)$ and $\gamma(t)$ for different perturbation strengths $k$, both in the coherent (solid lines) and incoherent case (dashed lines). }
    \label{fig:TwoFermions_fit}
\end{figure*}

In Fig.~\ref{fig:TwoFermions_fit}, we illustrate the time evolution of $|\nu(t)|$ and fit it using the same scaling function discussed in~\cite{paper_short} for single-particles. Panel (a) refers to the initial superposition state defined in \eqref{eq:TwoFermionState_menouno}, while panel (b) corresponds to the associated diagonal state, where all the off-diagonal terms are removed. Solid lines represent the numerical results, while dashed lines correspond to the fitted curves following the scaling law 
\begin{equation}\label{eq:LE_fit}
    \left| \nu(t) \right| \sim 1 - \beta(t) M^{\gamma(t)}. 
\end{equation}
The tables inside panels (a)-(b) report the fitted coefficients $\beta(t)$ and $\gamma(t)$, whose time-dependence is displayed in panels (c)-(d) of Fig.~\ref{fig:TwoFermions_fit}, for different values of the perturbation strength $k=1,10,100$. 
As in the other figures, the coherent (solid lines) and incoherent (dashed lines) scenarios are compared. 
While $\beta(t)$ remains nearly unchanged between the two cases, $\gamma(t)$ exhibits a sign change, becoming positive when quantum coherences are present. 
This behaviour highlights a coherence-enhanced orthogonalization effect, whereby quantum coherence accelerates the decay of $|\nu(t)|$ with increasing $M$. These results mirror the scaling law found in the single-particle case \cite{paper_short}, but here shaped by the anti-symmetrization of two fermions affected by a delta-perturbation. 

\section{Conclusions}\label{sec:conclusion}

In this paper, we studied the quantum dynamics and thermodynamics of a system of non-interacting fermions after an abrupt quench transformation, in which the fermions are coupled to a localized defect. As our main novel contribution, we considered the richer range of phenomena, including significant non-classical features, that emerge when the system is initialized in a coherent superposition of quantum states. To systematically explore these effects, we investigated a range of initial configurations, showing that the choice of the initial state plays a crucial role in determining both the degree of wave-function orthogonalization and the emergence of non-classical features during the post-quench evolution.

We started from the simplest case of a qubit, namely a superposition of two eigenstates of the unperturbed Hamiltonian, which we mapped on the Bloch sphere. In this setting, the post-quench response of the system is governed by the overlap amplitudes $\Lambda_{m,n}$ between the eigenstates of the unperturbed and perturbed Hamiltonians. In the strong perturbation regime ($k \rightarrow +\infty$), these overlaps exhibit a characteristic structure that approximates to a discretized hyperbola for large $n$. This gives rise to interference effects showcasing periodic singularities at integer multiples of $t = \pi$ in the LE.

Then, we extended our analysis to initial coherent states, which incorporate contributions from many energy levels with tunable amplitude and phase via a single complex parameter $\alpha$. In the strong perturbation regime, the LE displays a digital-like response, switching sharply between zero and one. This effect becomes more pronounced with increasing $\alpha$. From an energetic point of view, the average work performed by the defect on the system decreases and eventually saturates with increasing $|\alpha|$. However, if the coefficients of the initial superposition are modulated by the alternating phase factor $(-1)^{n/2}$, the average work remains constant.

In order to explore more structured forms of initial states comprising quantum coherence, we studied cat states, symmetric superpositions of coherent states with opposite phase-space displacements.
When the particle is initialized without any added phase factor, the LE remains close to unity for most times, with only brief and periodic drops, indicating practically no orthogonalization. The inclusion of the alternating phase factor $(-1)^{n/2}$, however, transforms the time-behaviour of LE into a series of sharp spikes recurring at $t = \pi$, making this the most efficient orthogonalizing configuration among all the ones considered. Initial cat state not only entails rapid and sustained suppression of the LE, but also yields the largest average work values, increasing consistently with $\alpha$. These findings identify the ``alternating-phase cat state'' as the optimal choice for inducing coherence-enhanced orthogonalization and maximizing energetic response to the quench transformation with a delta potential.

Finally, we explored the non-equilibrium response of two-fermions in order to quantitatively single-out how quantum correlations influence coherence-enhanced orthogonalization. We found that introducing coherences in the initial state of a single fermion yields dynamics very similar to those of two perturbed independent fermions subjected to the delta-perturbation. The average work increases with the superposition size only when the alternating phase factor $(-1)^{n/2}$ is present in the initial state, although it remains lower than in the coefficients of the initial superposition. The decay of the LE follows the same power-law determined for delta-perturbed single particles~\cite{paper_short}. Notably, the decay rate increases with the superposition size $M$, while it decreases when off-diagonal elements of the initial density operator are accounted for, confirming that quantum coherence enhances orthogonalization for both the single- and two-particle regimes. 

\subsection{Outlooks}

Among the main perspectives of this work is the study of a Fermi sea interacting with a single non-fully localized impurity trapped in an optical tweezer embedded within the gas~\cite{Hewitt2024quantum,CataniPRA2012,TylutkiRecatiPRA2017}. This would involve incorporating a general many-body description of the impurity atom via an appropriate field operator. 
Another interesting direction concerns to understand whether the introduction of a delta-like perturbation could generate entangled states in a Fermi gas, by employing optimal control techniques as e.g.~\cite{PoggiPRL2024}. Similarly, one could investigate the optimal conditions under which switching on the delta perturbation in a harmonic potential may lead to the implementation of thermodynamic cycles at maximum power. Among the thermodynamic cycles, one could consider quantum Otto heat engines like in Refs.~\cite{FogartyQST2020,BoubakourPRR2023}, the Feshbach engine~\cite{KellerPRR2020}, and the Stirling cycle as in Ref.~\cite{ChatterjeeAPS2021}.

\begin{acknowledgments}
B.D.~and S.G.~acknowledge support from the PNRR MUR project PE0000023-NQSTI funded by the European Union---Next Generation EU. G.D.C.~and S.G.~acknowledge financial support from and the Royal Society Project IES\textbackslash R3\textbackslash 223086 ``Dissipation-based quantum inference for out-of-equilibrium quantum many-body systems''. F.S.~acknowledges support from the European Union under the Horizon 2020 research and innovation programme (project OrbiDynaMIQs, GA No.~949438) and under the Horizon Europe program HORIZON-CL4-2022-QUANTUM-02-SGA (project PASQuanS2.1, GA no.~101113690).
\end{acknowledgments}

\appendix

\section{Analytical derivation of the overlap in the limit of strong perturbation \texorpdfstring{$k\rightarrow+\infty$}{TEXT}}\label{app:Appendix_strong_pert}

In the limit of strong interaction $k\rightarrow+\infty$, we can analytically calculate the overlap $\Lambda_{m,n} \equiv \braket{\psi_m'}{\psi_n}$ between the single-particle eigenfunctions of the perturbed and unperturbed Hamiltonians. As in the main text, the superscript indicates the perturbed case. Moreover, the only non-trivial case happens when both $m$ and $n$ are even numbers. Using the expressions given in \eqref{eq:unperturbed_psi} and \eqref{eq:perturbed_psi}, we have
\begin{equation}
\begin{aligned}
    &\Lambda_{m,n} = \braket{\psi_m'}{\psi_n} = \\ 
    &= \int_{-\infty}^{+\infty} dx \, (-1)^{m/2} \psi_m'^*(x) \, (-1)^{n/2} \psi_n(x) = \\
    &= (-1)^{\frac{m+n}{2}} \int_{-\infty}^{+\infty} dx \, \frac{1}{(2\pi)^{1/4}} \frac{1}{\sqrt{2^{m+1} (m+1)!}} e^{-\frac{\abs{x}^2}{4}} H_{m+1} \left( \frac{\abs{x}}{\sqrt{2}} \right)\times \\
    &\times\frac{1}{(2\pi)^{1/4}} \frac{1}{\sqrt{2^n n!}} e^{-x^2/4} H_n \left( \frac{x}{\sqrt{2}} \right) = \\
    &= \frac{(-1)^{\frac{m+n}{2}}}{\sqrt{\pi \, 2^{m+n+2} n!(m+1)!}} \int_{-\infty}^{+\infty} dx \, e^{-\frac{x^2}{2}} H_{m+1} \left( \frac{\abs{x}}{\sqrt{2}} \right) H_n \left( \frac{x}{\sqrt{2}} \right).
\end{aligned}    
\end{equation}
Calling $c_{m,n}$ the prefactor before the integral and changing the variable $y=x/\sqrt{2}$, we can develop the calculation
\begin{equation}\label{eq:app_Lambda_mn_begin}
\begin{aligned}    
    \Lambda_{m,n} &= c_{m,n} \sqrt{2}  \int_{-\infty}^{+\infty} dy \, e^{-y^2} H_{m+1} \left( \abs{y} \right) H_n \left( y \right) = \\
    &= c_{m,n} \sqrt{2} \left[  \int_{0}^{+\infty} dy \, e^{-y^2} H_{m+1} \left( y \right) H_n \left( y \right) + \right. \\
    &\left. +\int_{-\infty}^{0} dy \, e^{-y^2} H_{m+1} \left( -y \right) (-1)^n H_n \left( -y \right) \right] = \\
    &= c_{m,n} 2\sqrt{2}  \int_{0}^{+\infty} dy \, e^{-y^2} H_{m+1} \left( y \right) H_n \left( y \right),
\end{aligned}
\end{equation}
where we have used the symmetry of the Hermite polynomials $H_n(-x)=(-1)^nH_n(x)$. The Hermite polynomials with an even index can be expressed as
\begin{equation}\label{Hermite_even}
    H_n(x) = n! \sum_{l=0}^{n/2} \frac{(-1)^{\frac{n}{2}-l}}{(2l)!\left(\frac{n}{2}-l\right)!} (2x)^{2l} \,.
\end{equation}
Inserting this series inside the integral in \eqref{eq:app_Lambda_mn_begin}, we will need to calculate an integral of the kind (Eq.~(7), 376-3 of Ref.~\cite{Gradshteyn_Ryzhik_Zwillinger_Moll_2015})
\begin{equation}
\begin{aligned}
    &\int_0^{+\infty} dx \, e^{-2\alpha x^2} H_{2n+1}(x)x^{\nu} = \\
    &=(-1)^{n} 2^{2n-\nu/2} \frac{\Gamma\left(\frac{\nu+1}{2} \right) \Gamma\left(n+\frac{1}{2}\right)}{\sqrt{\pi}\alpha^{\nu/2+1}} F\left( -n,\frac{\nu}{2}+1;\frac{3}{2};\frac{1}{2\alpha} \right),
\end{aligned}    
\end{equation}
where $F(\alpha,\beta;\gamma;z)=_2F_1(\alpha,\beta;\gamma;z)$ is the Gauss hypergeometric function. This integral reduces to our case with $\nu=2k$, $\alpha=1/2$ and $n=m/2$
\begin{equation}
\begin{aligned}
    &\int_0^{+\infty} dx \, e^{-x^2} H_{m+1}(x)x^{2k} =\\ 
    & = (-1)^{m/2} 2^{m-k} \frac{\Gamma(1+k) \Gamma\left(\frac{m+3}{2}\right)}{\sqrt{\pi}\left(\frac{1}{2}\right)^{k+1}} F\left( -\frac{m}{2},k+1;\frac{3}{2};1 \right).
\end{aligned}
\end{equation}
In the case where $\text{Re}(\gamma)>\text{Re}(\alpha+\beta)$, we know (Eq.~(9), 122-1 of \cite{Gradshteyn_Ryzhik_Zwillinger_Moll_2015}) that
\begin{equation}
    F(\alpha,\beta;\gamma;1) = \frac{\Gamma(\gamma) \Gamma(\gamma-\alpha-\beta)}{\Gamma(\gamma-\alpha)\Gamma(\gamma-\beta)} \,,
\end{equation}
thus in our case, imposing $m>2k-1$, 
we have
\begin{equation}
    F\left( -\frac{m}{2},k+1;\frac{3}{2};1 \right) = \frac{\Gamma\left(\frac{3}{2}\right) \Gamma\left(\frac{3}{2}+\frac{m}{2}-(k+1)\right)}{\Gamma\left(\frac{m+3}{2}\right)\Gamma\left(\frac{3}{2}-(k+1)\right)} \,.
\end{equation}
The integral is finally
\begin{equation}\label{final_integral_even}
    \int_0^{+\infty} dx \, e^{-x^2} H_{m+1}(x)x^{2k} = (-1)^{m/2} 2^m k! \frac{\Gamma\left( \frac{m+1}{2}-k\right)}{\Gamma\left(\frac{1}{2}-k \right)} \,.
\end{equation}
Accordingly, the generic overlap \eqref{eq:app_Lambda_mn_begin} can be calculated exactly
\begin{equation}
\begin{aligned}    
    \Lambda_{m,n} &= c_{m,n} 2\sqrt{2} n! \sum_{l=0}^{n/2} \frac{(-1)^{\frac{n}{2}-l} 2^{2l}}{(2l)!\left(\frac{n}{2}-l\right)!} \int_{0}^{+\infty} dy \, e^{-y^2} H_{m+1} \left( y \right)  y^{2l} \\
    &= \frac{(-1)^{\frac{m+n}{2}}}{\sqrt{(2\pi) 2^{m+n+1} n!(m+1)!}} 2\sqrt{2} n! \sum_{l=0}^{n/2} \frac{(-1)^{\frac{n}{2}-l} 2^{2l}}{(2l)!\left(\frac{n}{2}-l\right)!}\times \\
    &\times (-1)^{m/2} 2^m l! \frac{\Gamma\left( \frac{m+1}{2}-l\right)}{\Gamma\left(\frac{1}{2}-l \right)} = \\
    &= \frac{1}{\sqrt{\pi}} \sqrt{\frac{2^{m+1}}{(m+1)!}} \sqrt{\frac{n!}{2^n}} \sum_{l=0}^{n/2} \frac{(-1)^{-l} 2^{2l} l!}{(2l)!\left(\frac{n}{2}-l\right)!} \frac{\Gamma\left( \frac{m+1}{2}-l\right)}{\Gamma\left(\frac{1}{2}-l \right)} \\
    &= \sqrt{\frac{2^{m+1}}{(m+1)!}} \sqrt{\frac{n!}{2^n}} \sum_{l=0}^{n/2} \frac{(-1)^{l}}{\left(\frac{n}{2}-l\right)!} \frac{\Gamma\left( \frac{m+1}{2}-l\right)}{\pi (-1)^l} = \\
    &=\frac{1}{\pi} \sqrt{\frac{2^{m+1}}{(m+1)!}} \sqrt{\frac{n!}{2^n}} \sum_{l=0}^{n/2} \frac{\Gamma\left( \frac{m+1}{2}-l\right)}{\Gamma\left( \frac{n}{2}-l+1\right)} \,,
\end{aligned}
\end{equation}
where we have used the identities 
\begin{align}
    \frac{2^{2n}n!}{(2n)!} &= \frac{\sqrt{\pi}}{\Gamma\left( n+\frac{1}{2} \right)} \\
    \Gamma\left( n+\frac{1}{2} \right) \Gamma\left( \frac{1}{2}-n \right) &= \pi \sec{\pi n} = \pi (-1)^n \,.
\end{align}
The summation converges so that
\begin{equation}
\begin{aligned}
    &\Lambda_{m,n} = \frac{1}{\pi} \sqrt{\frac{2^{m+1}}{(m+1)!}} \sqrt{\frac{n!}{2^n}} \frac{m+1}{m+1-n} \frac{\Gamma\left( \frac{m+1}{2}\right)}{\Gamma\left( \frac{n}{2}+1 \right)} = \\
    &= \Lambda_{m,0} \sqrt{\frac{n!}{2^n}} \frac{m+1}{m+1-n} \frac{\sqrt{2^n}}{n!!} = \Lambda_{m,0} \sqrt{\frac{(n-1)!!}{n!!}} \frac{m+1}{m+1-n} \,,
\end{aligned}
\end{equation}
where we gathered the overlap with the unperturbed ground state:
\begin{equation}
    \Lambda_{m0} = \sqrt{\frac{2^{m+1}}{(m+1)!}} \frac{\Gamma \left(\frac{m+1}{2} \right)}{\pi} \,.
\end{equation}
It is now straightforward to derive the recursive relation 
\begin{equation}
    \frac{\Lambda_{m,n}}{\Lambda_{m,n-2}} = \sqrt{\frac{n-1}{n}} \frac{m-n+3}{m-n+1} \,.
\end{equation}

\bibliography{OrthogonalityCatastrophe.bib}

\begin{thebibliography}{48}
\providecommand{\natexlab}[1]{#1}
\providecommand{\url}[1]{\texttt{#1}}
\expandafter\ifx\csname urlstyle\endcsname\relax
  \providecommand{\doi}[1]{doi: #1}\else
  \providecommand{\doi}{doi: \begingroup \urlstyle{rm}\Url}\fi

\bibitem[Anderson(1967)]{AndersonPRL1967}
P.~W. Anderson.
\newblock {Infrared Catastrophe in Fermi Gases with Local Scattering Potentials}.
\newblock \emph{Phys. Rev. Lett.}, 18:\penalty0 1049--1051, Jun 1967.
\newblock \doi{10.1103/PhysRevLett.18.1049}.
\newblock URL \url{https://link.aps.org/doi/10.1103/PhysRevLett.18.1049}.

\bibitem[Mahan(1967)]{MahanPR1967}
G.~D. Mahan.
\newblock {Excitons in Metals: Infinite Hole Mass}.
\newblock \emph{Phys. Rev.}, 163:\penalty0 612--617, Nov 1967.
\newblock \doi{10.1103/PhysRev.163.612}.
\newblock URL \url{https://link.aps.org/doi/10.1103/PhysRev.163.612}.

\bibitem[Knap et~al.(2012)Knap, Shashi, Nishida, Imambekov, Abanin, and Demler]{Knap2012}
M.~Knap, A.~Shashi, Y.~Nishida, A.~Imambekov, D.~A. Abanin, and E.~Demler.
\newblock {Time-Dependent Impurity in Ultracold Fermions: Orthogonality Catastrophe and Beyond}.
\newblock \emph{Phys. Rev. X}, 2:\penalty0 041020, Dec 2012.
\newblock \doi{10.1103/PhysRevX.2.041020}.
\newblock URL \url{https://link.aps.org/doi/10.1103/PhysRevX.2.041020}.

\bibitem[Cetina et~al.(2016)Cetina, Jag, Lous, Fritsche, Walraven, Grimm, Levinsen, Parish, Schmidt, Knap, and Demler]{Cetina2016}
Marko Cetina, Michael Jag, Rianne~S. Lous, Isabella Fritsche, Jook T.~M. Walraven, Rudolf Grimm, Jesper Levinsen, Meera~M. Parish, Richard Schmidt, Michael Knap, and Eugene Demler.
\newblock Ultrafast many-body interferometry of impurities coupled to a fermi sea.
\newblock \emph{Science}, 354\penalty0 (6308):\penalty0 96–99, October 2016.
\newblock ISSN 1095-9203.
\newblock \doi{10.1126/science.aaf5134}.
\newblock URL \url{http://dx.doi.org/10.1126/science.aaf5134}.

\bibitem[Peres(1984)]{PeresPRA84}
A.~Peres.
\newblock Stability of quantum motion in chaotic and regular systems.
\newblock \emph{Phys. Rev. A}, 30:\penalty0 1610--1615, Oct 1984.
\newblock \doi{10.1103/PhysRevA.30.1610}.
\newblock URL \url{https://link.aps.org/doi/10.1103/PhysRevA.30.1610}.

\bibitem[Jalabert and Pastawski(2001)]{JalabertPRL2001}
R.~A. Jalabert and H.~M. Pastawski.
\newblock {Environment-Independent Decoherence Rate in Classically Chaotic Systems}.
\newblock \emph{Phys. Rev. Lett.}, 86:\penalty0 2490--2493, Mar 2001.
\newblock \doi{10.1103/PhysRevLett.86.2490}.
\newblock URL \url{https://link.aps.org/doi/10.1103/PhysRevLett.86.2490}.

\bibitem[Cucchietti et~al.(2003)Cucchietti, Dalvit, Paz, and Zurek]{CucchiettiPRL2003}
F.~M. Cucchietti, D.~A.~R. Dalvit, J.~P. Paz, and W.~H. Zurek.
\newblock {Decoherence and the Loschmidt Echo}.
\newblock \emph{Phys. Rev. Lett.}, 91:\penalty0 210403, Nov 2003.
\newblock \doi{10.1103/PhysRevLett.91.210403}.
\newblock URL \url{https://link.aps.org/doi/10.1103/PhysRevLett.91.210403}.

\bibitem[Silva(2008)]{SilvaPRL2008}
A.~Silva.
\newblock {Statistics of the Work Done on a Quantum Critical System by Quenching a Control Parameter}.
\newblock \emph{Phys. Rev. Lett.}, 101:\penalty0 120603, Sep 2008.
\newblock \doi{10.1103/PhysRevLett.101.120603}.
\newblock URL \url{https://link.aps.org/doi/10.1103/PhysRevLett.101.120603}.

\bibitem[Lostaglio et~al.(2023)Lostaglio, Belenchia, Levy, Hern{\'{a}}ndez-G{\'{o}}mez, Fabbri, and Gherardini]{LostaglioQuantum2023}
M.~Lostaglio, A.~Belenchia, A.~Levy, S.~Hern{\'{a}}ndez-G{\'{o}}mez, N.~Fabbri, and S.~Gherardini.
\newblock Kirkwood-{D}irac quasiprobability approach to the statistics of incompatible observables.
\newblock \emph{{Quantum}}, 7:\penalty0 1128, October 2023.
\newblock ISSN 2521-327X.
\newblock \doi{10.22331/q-2023-10-09-1128}.
\newblock URL \url{https://doi.org/10.22331/q-2023-10-09-1128}.

\bibitem[Goold et~al.(2011)Goold, Fogarty, Lo~Gullo, Paternostro, and Busch]{goold_orthogonality_2011}
J.~Goold, T.~Fogarty, N.~Lo~Gullo, M.~Paternostro, and Th. Busch.
\newblock Orthogonality catastrophe as a consequence of qubit embedding in an ultracold {F}ermi gas.
\newblock \emph{Phys. Rev. A}, 84:\penalty0 063632, Dec 2011.
\newblock \doi{10.1103/PhysRevA.84.063632}.
\newblock URL \url{https://link.aps.org/doi/10.1103/PhysRevA.84.063632}.

\bibitem[Nozi\`eres and De~Dominicis(1969)]{NozieresPR1969}
P.~Nozi\`eres and C.~T. De~Dominicis.
\newblock {Singularities in the X-Ray Absorption and Emission of Metals. III. One-Body Theory Exact Solution}.
\newblock \emph{Phys. Rev.}, 178:\penalty0 1097--1107, Feb 1969.
\newblock \doi{10.1103/PhysRev.178.1097}.
\newblock URL \url{https://link.aps.org/doi/10.1103/PhysRev.178.1097}.

\bibitem[Fusco et~al.(2014)Fusco, Pigeon, Apollaro, Xuereb, Mazzola, Campisi, Ferraro, Paternostro, and De~Chiara]{Fusco2014}
L.~Fusco, S.~Pigeon, T.~J.~G. Apollaro, A.~Xuereb, L.~Mazzola, M.~Campisi, A.~Ferraro, M.~Paternostro, and G.~De~Chiara.
\newblock {Assessing the Nonequilibrium Thermodynamics in a Quenched Quantum Many-Body System via Single Projective Measurements}.
\newblock \emph{Phys. Rev. X}, 4:\penalty0 031029, Aug 2014.
\newblock \doi{10.1103/PhysRevX.4.031029}.
\newblock URL \url{https://link.aps.org/doi/10.1103/PhysRevX.4.031029}.

\bibitem[Alba and Calabrese(2017)]{AlbaPNAS2017}
Vincenzo Alba and Pasquale Calabrese.
\newblock {Entanglement and thermodynamics after a quantum quench in integrable systems}.
\newblock \emph{PNAS}, 114:\penalty0 7947--7951, 2017.
\newblock \doi{10.1073/pnas.1703516114}.
\newblock URL \url{https://www.pnas.org/doi/10.1073/pnas.1703516114}.

\bibitem[Santini et~al.(2023)Santini, Solfanelli, Gherardini, and Collura]{santini_work_2023}
A.~Santini, A.~Solfanelli, S.~Gherardini, and M.~Collura.
\newblock Work statistics, quantum signatures, and enhanced work extraction in quadratic fermionic models.
\newblock \emph{Phys. Rev. B}, 108\penalty0 (10):\penalty0 104308, September 2023.
\newblock ISSN 2469-9950, 2469-9969.
\newblock \doi{10.1103/PhysRevB.108.104308}.
\newblock URL \url{https://link.aps.org/doi/10.1103/PhysRevB.108.104308}.

\bibitem[Gherardini and De~Chiara(2024)]{GherardiniTutorial}
S.~Gherardini and G.~De~Chiara.
\newblock {Quasiprobabilities in Quantum Thermodynamics and Many-Body Systems}.
\newblock \emph{PRX Quantum}, 5:\penalty0 030201, Sep 2024.
\newblock \doi{10.1103/PRXQuantum.5.030201}.
\newblock URL \url{https://link.aps.org/doi/10.1103/PRXQuantum.5.030201}.

\bibitem[Yoshimura and S\'a(2025)]{YoshimuraPRE2025}
Takato Yoshimura and Lucas S\'a.
\newblock Theory of irreversibility in quantum many-body systems.
\newblock \emph{Phys. Rev. E}, 111:\penalty0 064135, Jun 2025.
\newblock \doi{10.1103/82f6-qdyd}.
\newblock URL \url{https://link.aps.org/doi/10.1103/82f6-qdyd}.

\bibitem[Jaksch and Zoller(2005)]{Jaksch2005}
D.~Jaksch and P.~Zoller.
\newblock The cold atom hubbard toolbox.
\newblock \emph{Ann. Phys.}, 315\penalty0 (1):\penalty0 52--79, 2005.
\newblock \doi{https://doi.org/10.1016/j.aop.2004.09.010}.
\newblock URL \url{https://www.sciencedirect.com/science/article/pii/S0003491604001782}.

\bibitem[Recati et~al.(2005)Recati, Fedichev, Zwerger, von Delft, and Zoller]{Recati2005}
A.~Recati, P.~O. Fedichev, W.~Zwerger, J.~von Delft, and P.~Zoller.
\newblock Atomic quantum dots coupled to a reservoir of a superfluid bose-einstein condensate.
\newblock \emph{Phys. Rev. Lett.}, 94:\penalty0 040404, Feb 2005.
\newblock \doi{10.1103/PhysRevLett.94.040404}.
\newblock URL \url{https://link.aps.org/doi/10.1103/PhysRevLett.94.040404}.

\bibitem[Donelli et~al.(2025)Donelli, De~Chiara, Scazza, and Gherardini]{paper_short}
B.~Donelli, G.~De~Chiara, F.~Scazza, and S.~Gherardini.
\newblock Orthogonalization speed-up from quantum coherence after a sudden quench.
\newblock \emph{arXiv:2507.21313}, 2025.
\newblock URL \url{https://doi.org/10.48550/arXiv.2507.21313}.

\bibitem[Yunger~Halpern et~al.(2018)Yunger~Halpern, Swingle, and Dressel]{yunger2018quasiprobability}
N.~Yunger~Halpern, B.~Swingle, and J.~Dressel.
\newblock Quasiprobability behind the out-of-time-ordered correlator.
\newblock \emph{Phys. Rev. A}, 97:\penalty0 042105, Apr 2018.
\newblock \doi{10.1103/PhysRevA.97.042105}.
\newblock URL \url{https://link.aps.org/doi/10.1103/PhysRevA.97.042105}.

\bibitem[Arvidsson-Shukur et~al.(2021)Arvidsson-Shukur, Chevalier~Drori, and Yunger~Halpern]{ArvidssonShukurJPA2021}
D.~R.~M. Arvidsson-Shukur, J.~Chevalier~Drori, and N.~Yunger~Halpern.
\newblock Conditions tighter than noncommutation needed for nonclassicality.
\newblock \emph{J. Phys. A: Math. Theor.}, 54:\penalty0 284001, 2021.
\newblock \doi{10.1088/1751-8121/ac0289}.
\newblock URL \url{https://iopscience.iop.org/article/10.1088/1751-8121/ac0289}.

\bibitem[De~Bi\`evre(2021)]{DeBievrePRL2021}
S.~De~Bi\`evre.
\newblock {Complete Incompatibility, Support Uncertainty, and Kirkwood-Dirac Nonclassicality}.
\newblock \emph{Phys. Rev. Lett.}, 127:\penalty0 190404, Nov 2021.
\newblock \doi{10.1103/PhysRevLett.127.190404}.
\newblock URL \url{https://link.aps.org/doi/10.1103/PhysRevLett.127.190404}.

\bibitem[Budiyono and Dipojono(2023)]{BudiyonoPRAquantifying}
A.~Budiyono and H.~K. Dipojono.
\newblock {Quantifying quantum coherence via Kirkwood-Dirac quasiprobability}.
\newblock \emph{Phys. Rev. A}, 107:\penalty0 022408, Feb 2023.
\newblock \doi{10.1103/PhysRevA.107.022408}.
\newblock URL \url{https://link.aps.org/doi/10.1103/PhysRevA.107.022408}.

\bibitem[Wagner et~al.(2024)Wagner, Schwartzman-Nowik, Paiva, Te'eni, Ruiz-Molero, Barbosa, Cohen, and Galv{\~a}o]{wagner2023quantum}
R.~Wagner, Z.~Schwartzman-Nowik, I.~L. Paiva, A.~Te'eni, A.~Ruiz-Molero, R.~Soares Barbosa, E.~Cohen, and E.~F. Galv{\~a}o.
\newblock {Quantum circuits for measuring weak values, Kirkwood--Dirac quasiprobability distributions, and state spectra}.
\newblock \emph{Quantum Sci. Technol.}, 9:\penalty0 015030, 2024.
\newblock \doi{10.1088/2058-9565/ad124c}.
\newblock URL \url{https://iopscience.iop.org/article/10.1088/2058-9565/ad124c}.

\bibitem[Arvidsson-Shukur et~al.(2024)Arvidsson-Shukur, Braasch~Jr., De~Bi\`evre, Dressel, Jordan, Langrenez, Lostaglio, Lundeen, and Yunger~Halpern]{ArvidssonShukur2024review}
D.~R.~M. Arvidsson-Shukur, W.~F. Braasch~Jr., S.~De~Bi\`evre, J.~Dressel, A.~N. Jordan, C.~Langrenez, M.~Lostaglio, J.~S. Lundeen, and N.~Yunger~Halpern.
\newblock {Properties and Applications of the Kirkwood-Dirac Distribution}.
\newblock \emph{New J. Phys.}, 26:\penalty0 121201, 2024.
\newblock \doi{10.1088/1367-2630/ada05d}.
\newblock URL \url{https://iopscience.iop.org/article/10.1088/1367-2630/ada05d}.

\bibitem[Hern{\'a}ndez-G{\'o}mez et~al.(2024)Hern{\'a}ndez-G{\'o}mez, Isogawa, Belenchia, Levy, Fabbri, Gherardini, and Cappellaro]{hernandez2024Interfero}
S.~Hern{\'a}ndez-G{\'o}mez, T.~Isogawa, A.~Belenchia, A.~Levy, N.~Fabbri, S.~Gherardini, and P.~Cappellaro.
\newblock Interferometry of quantum correlation functions to access quasiprobability distribution of work.
\newblock \emph{Npj Quantum Inf.}, 10:\penalty0 115, 2024.
\newblock \doi{10.1038/s41534-024-00913-x}.
\newblock URL \url{https://www.nature.com/articles/s41534-024-00913-x}.

\bibitem[Fogarty et~al.(2020)Fogarty, Deffner, Busch, and Campbell]{Fogarty2020}
T.~Fogarty, S.~Deffner, Th. Busch, and S.~Campbell.
\newblock {Orthogonality Catastrophe as a Consequence of the Quantum Speed Limit}.
\newblock \emph{Phys. Rev. Lett.}, 124:\penalty0 110601, Mar 2020.
\newblock \doi{10.1103/PhysRevLett.124.110601}.
\newblock URL \url{https://link.aps.org/doi/10.1103/PhysRevLett.124.110601}.

\bibitem[Busch and Huyet(2003)]{busch_low-density_2003}
Th. Busch and G.~Huyet.
\newblock Low-density, one-dimensional quantum gases in a split trap.
\newblock \emph{J. Phys. B: At. Mol. Opt. Phys.}, 36\penalty0 (12):\penalty0 2553--2562, June 2003.
\newblock ISSN 0953-4075, 1361-6455.
\newblock \doi{10.1088/0953-4075/36/12/313}.
\newblock URL \url{https://iopscience.iop.org/article/10.1088/0953-4075/36/12/313}.

\bibitem[Wigner(1932)]{Wigner1932}
E.~Wigner.
\newblock On the quantum correction for thermodynamic equilibrium.
\newblock \emph{Physical Review}, 40\penalty0 (5):\penalty0 749–759, June 1932.
\newblock ISSN 0031-899X.
\newblock \doi{10.1103/physrev.40.749}.
\newblock URL \url{http://dx.doi.org/10.1103/PhysRev.40.749}.

\bibitem[Gerry and Knight(2004)]{Gerry2004}
C.~Gerry and P.~Knight.
\newblock \emph{Introductory Quantum Optics}.
\newblock Cambridge University Press, October 2004.
\newblock ISBN 9780511791239.
\newblock \doi{10.1017/cbo9780511791239}.
\newblock URL \url{http://dx.doi.org/10.1017/CBO9780511791239}.

\bibitem[Loudon(2000)]{Loudon2000}
R.~Loudon.
\newblock \emph{{The Quantum Theory of Light}}.
\newblock Oxford University PressOxford, September 2000.
\newblock ISBN 9781383020151.
\newblock \doi{10.1093/oso/9780198501770.001.0001}.
\newblock URL \url{http://dx.doi.org/10.1093/oso/9780198501770.001.0001}.

\bibitem[Campisi et~al.(2011)Campisi, H\"{a}nggi, and Talkner]{CampisiRMP2011}
M.~Campisi, P.~H\"{a}nggi, and P.~Talkner.
\newblock Colloquium: Quantum fluctuation relations: Foundations and applications.
\newblock \emph{Reviews of Modern Physics}, 83\penalty0 (3):\penalty0 771–791, July 2011.
\newblock ISSN 1539-0756.
\newblock \doi{10.1103/revmodphys.83.771}.
\newblock URL \url{http://dx.doi.org/10.1103/RevModPhys.83.771}.

\bibitem[Bratteli and Robinson(1997)]{Bratteli1997}
O.~Bratteli and D.~W. Robinson.
\newblock \emph{Operator Algebras and Quantum Statistical Mechanics}.
\newblock Springer Berlin Heidelberg, 1997.
\newblock ISBN 9783662034446.
\newblock \doi{10.1007/978-3-662-03444-6}.
\newblock URL \url{http://dx.doi.org/10.1007/978-3-662-03444-6}.

\bibitem[Margenau and Hill(1961)]{margenau1961correlation}
H.~Margenau and R.~N. Hill.
\newblock Correlation between measurements in quantum theory.
\newblock \emph{Prog. Theor. Phys.}, 26\penalty0 (5):\penalty0 722--738, 1961.
\newblock \doi{10.1143/PTP.26.722}.

\bibitem[Allahverdyan(2014)]{allahverdyan2014nonequilibrium}
A.~E. Allahverdyan.
\newblock Nonequilibrium quantum fluctuations of work.
\newblock \emph{Phys. Rev. E}, 90:\penalty0 032137, Sep 2014.
\newblock \doi{10.1103/PhysRevE.90.032137}.
\newblock URL \url{https://link.aps.org/doi/10.1103/PhysRevE.90.032137}.

\bibitem[Lostaglio(2018)]{lostaglio2018quantum}
M.~Lostaglio.
\newblock {Quantum Fluctuation Theorems, Contextuality, and Work Quasiprobabilities}.
\newblock \emph{Phys. Rev. Lett.}, 120:\penalty0 040602, Jan 2018.
\newblock \doi{10.1103/PhysRevLett.120.040602}.
\newblock URL \url{https://link.aps.org/doi/10.1103/PhysRevLett.120.040602}.

\bibitem[D{\'\i}az et~al.(2020)D{\'\i}az, Guarnieri, and Paternostro]{diaz2020quantum}
M.~G. D{\'\i}az, G.~Guarnieri, and M.~Paternostro.
\newblock {Quantum Work Statistics with Initial Coherence}.
\newblock \emph{Entropy}, 22\penalty0 (11):\penalty0 1223, 2020.
\newblock \doi{10.3390/e22111223}.
\newblock URL \url{https://www.mdpi.com/1099-4300/22/11/1223}.

\bibitem[Hern\'andez-G\'omez et~al.(2024)Hern\'andez-G\'omez, Gherardini, Belenchia, Lostaglio, Levy, and Fabbri]{hernandez2022experimental}
S.~Hern\'andez-G\'omez, S.~Gherardini, A.~Belenchia, M.~Lostaglio, A.~Levy, and N.~Fabbri.
\newblock Projective measurements can probe nonclassical work extraction and time correlations.
\newblock \emph{Phys. Rev. Res.}, 6:\penalty0 023280, Jun 2024.
\newblock \doi{10.1103/PhysRevResearch.6.023280}.
\newblock URL \url{https://link.aps.org/doi/10.1103/PhysRevResearch.6.023280}.

\bibitem[Pei et~al.(2023)Pei, Chen, and Quan]{PeiPRE2023}
J.-H. Pei, J.-F. Chen, and H.~T. Quan.
\newblock {Exploring quasiprobability approaches to quantum work in the presence of initial coherence: Advantages of the Margenau-Hill distribution}.
\newblock \emph{Phys. Rev. E}, 108:\penalty0 054109, Nov 2023.
\newblock \doi{10.1103/PhysRevE.108.054109}.
\newblock URL \url{https://link.aps.org/doi/10.1103/PhysRevE.108.054109}.

\bibitem[Hewitt et~al.(2024)Hewitt, Bertheas, Jain, Nishida, and Barontini]{Hewitt2024quantum}
T.~Hewitt, T.~Bertheas, M.~Jain, Y.~Nishida, and G.~Barontini.
\newblock {Controlling the interactions in a cold atom quantum impurity system}.
\newblock \emph{Quantum Sci. Technol.}, 9:\penalty0 035039, 2024.
\newblock \doi{10.1088/2058-9565/ad4c91}.
\newblock URL \url{https://iopscience.iop.org/article/10.1088/2058-9565/ad4c91}.

\bibitem[Catani et~al.(2012)Catani, Lamporesi, Naik, Gring, Inguscio, Minardi, Kantian, and Giamarchi]{CataniPRA2012}
J.~Catani, G.~Lamporesi, D.~Naik, M.~Gring, M.~Inguscio, F.~Minardi, A.~Kantian, and T.~Giamarchi.
\newblock Quantum dynamics of impurities in a one-dimensional bose gas.
\newblock \emph{Phys. Rev. A}, 85:\penalty0 023623, Feb 2012.
\newblock \doi{10.1103/PhysRevA.85.023623}.
\newblock URL \url{https://link.aps.org/doi/10.1103/PhysRevA.85.023623}.

\bibitem[Tylutki et~al.(2017)Tylutki, Astrakharchik, and Recati]{TylutkiRecatiPRA2017}
Marek Tylutki, G.~E. Astrakharchik, and Alessio Recati.
\newblock Coherent oscillations in small fermi-polaron systems.
\newblock \emph{Phys. Rev. A}, 96:\penalty0 063603, Dec 2017.
\newblock \doi{10.1103/PhysRevA.96.063603}.
\newblock URL \url{https://link.aps.org/doi/10.1103/PhysRevA.96.063603}.

\bibitem[Poggi et~al.(2024)Poggi, De~Chiara, Campbell, and Kiely]{PoggiPRL2024}
P.~M. Poggi, G.~De~Chiara, S.~Campbell, and A.~Kiely.
\newblock {Universally Robust Quantum Control}.
\newblock \emph{Phys. Rev. Lett.}, 132:\penalty0 193801, May 2024.
\newblock \doi{10.1103/PhysRevLett.132.193801}.
\newblock URL \url{https://link.aps.org/doi/10.1103/PhysRevLett.132.193801}.

\bibitem[Fogarty and Busch(2020)]{FogartyQST2020}
Thomás Fogarty and Thomas Busch.
\newblock A many-body heat engine at criticality.
\newblock \emph{Quantum Sci. Technol.}, 6\penalty0 (1):\penalty0 015003, 2020.
\newblock ISSN 2058-9565.
\newblock \doi{10.1088/2058-9565/abbc63}.
\newblock URL \url{http://dx.doi.org/10.1088/2058-9565/abbc63}.

\bibitem[Boubakour et~al.(2023)Boubakour, Fogarty, and Busch]{BoubakourPRR2023}
Mohamed Boubakour, Thom\'as Fogarty, and Thomas Busch.
\newblock Interaction-enhanced quantum heat engine.
\newblock \emph{Phys. Rev. Res.}, 5:\penalty0 013088, Feb 2023.
\newblock \doi{10.1103/PhysRevResearch.5.013088}.
\newblock URL \url{https://link.aps.org/doi/10.1103/PhysRevResearch.5.013088}.

\bibitem[Keller et~al.(2020)Keller, Fogarty, Li, and Busch]{KellerPRR2020}
Tim Keller, Thom\'as Fogarty, Jing Li, and Thomas Busch.
\newblock {Feshbach engine in the Thomas-Fermi regime}.
\newblock \emph{Phys. Rev. Res.}, 2:\penalty0 033335, Aug 2020.
\newblock \doi{10.1103/PhysRevResearch.2.033335}.
\newblock URL \url{https://link.aps.org/doi/10.1103/PhysRevResearch.2.033335}.

\bibitem[Chatterjee et~al.(2021)Chatterjee, Koner, Chatterjee, and Kumar]{ChatterjeeAPS2021}
S.~Chatterjee, A.~Koner, S.~Chatterjee, and C.~Kumar.
\newblock {Temperature-dependent maximization of work and efficiency in a degeneracy-assisted quantum Stirling heat engine}.
\newblock \emph{Phys. Rev. E}, 103:\penalty0 062109, Jun 2021.
\newblock \doi{10.1103/PhysRevE.103.062109}.
\newblock URL \url{https://link.aps.org/doi/10.1103/PhysRevE.103.062109}.

\bibitem[Gradshteyn et~al.(2015)Gradshteyn, Ryzhik, Zwillinger, and Moll]{Gradshteyn_Ryzhik_Zwillinger_Moll_2015}
I.~S. Gradshteyn, I.~M. Ryzhik, D.~Zwillinger, and V.~H. Moll.
\newblock \emph{Table of integrals, series, and products}.
\newblock Academic Press, 2015.

\end{thebibliography}

\end{document}